\documentclass[preprint,10pt,3p]{elsarticle}
\usepackage{amsmath,amssymb}
\usepackage{graphicx}
\usepackage{multicol}
\usepackage{epstopdf}
\usepackage{subfig}

\usepackage{setspace}
\journal{Chaos, Solitons and Fractals}

\newproof{proof}{Proof}

\newcommand{\pdI}[2]{\ensuremath{\frac{\partial{#1}}{\partial{#2}}}}
\newcommand{\ddI}[2]{\ensuremath{\frac{d{#1}}{d{#2}}}}
\newcommand{\la}{\left\langle}
\newcommand{\ra}{\right\rangle}

\begin{document}

\begin{frontmatter}

\title{Least Squares Shadowing Sensitivity Analysis of a Modified Kuramoto-Sivashinsky Equation}

\author[mit]{Patrick J. Blonigan\corref{cor}}
\ead{blonigan@mit.edu}

\author[mit]{Qiqi Wang}
\ead{qiqi@mit.edu}

\address[mit]{Department of Aeronautics and Astronautics,
Massachusetts Institute of Technology, 77 Massachusetts Ave, Cambridge,
MA 02139, United States}

\begin{keyword}
Sensitivity Analysis \sep Kuramoto-Sivashinsky Equation
\end{keyword}

\begin{abstract}
Computational methods for sensitivity analysis are invaluable tools for scientists and engineers investigating a wide range of physical phenomena.  However, many of these methods fail when applied to chaotic systems, such as the Kuramoto-Sivashinsky (K-S) equation, which models a number of different chaotic systems found in nature.  The following paper discusses the application of a new sensitivity analysis method developed by the authors to a modified K-S equation.  We find that least squares shadowing sensitivity analysis computes accurate gradients for solutions corresponding to a wide range of system parameters.  
\end{abstract}

\end{frontmatter}

\section{Introduction}

Sensitivity analysis is of great importance to scientists and engineers.  It is used to compute sensitivity
derivatives of key quantities of interest to parameters that influence
a system which is governed by some ordinary differential equation (ODE) or partial differential equation (PDE).  These sensitivity derivatives can be used for design optimization, inverse problems,
data assimilation, and uncertainty quantification.

Sensitivity analysis for chaotic dynamical systems is important because of the prevalence of chaos in many scientific and engineering fields. A relatively simple PDE that can be chaotic, the Kuramoto-Sivashinsky (K-S) equation, \eqref{e:KS}, is a great illustration of how widespread chaos is in nature, as it has been found to model a wide range of physical phenomena.  

\begin{equation}
\pdI{u}{t} = - u \pdI{u}{x} - \pdI{^2 u}{x^2} - \pdI{^4 u}{x^4} 
\label{e:KS}
\end{equation}

Kuramoto derived the equation for angular-phase turbulence for a system of reaction-diffusion equations modeling the Belouzov-Zabotinskii reaction in three spatial dimensions \cite{Kuramoto:1976:reaction,Kuramoto:1978:reaction}. Sivashinsky also derived the equation to model the evolution of instabilities in a distributed plane flame front \cite{Sivashinsky:1977:flames1,Sivashinsky:1977:flames2}. In addition the K-S equation has also been shown to be a model of Poiseuille flow of a film layer on an inclined plane \cite{Sivashinsky:1980:film}. 

In the study of periodic and chaotic systems, long time averaged quantities, such as mean temperature and mean aerodynamic forces of turbulent fluid flows, are of interest.  However, many sensitivity analysis methods fail when applied to long time averaged quantities in chaotic dynamical systems. The sensitivity gradient predicted by methods including the adjoint method are observed to diverge as simulation time is increased \cite{Lea:2000:climate_sens}.  

A recently developed method, the least squares shadowing (LSS) method, can compute accurate gradients for ergodic chaotic systems \cite{Wang:2013:LSS1,Wang:2013:LSS2}.  The LSS method finds a perturbed trajectory (or solution) that does not diverge exponentially from some trajectory in phase space.  This non-diverging trajectory, called a ``shadow trajectory'', has its existence guaranteed by the shadowing lemma \cite{Pilyugin:1999:shadow} for a large number of chaotic systems and can be used to compute sensitivities.  

This paper discusses the application of the LSS method to a one-dimensional modified K-S equation:

\begin{gather}
\pdI{u}{t} = - (u + c) \pdI{u}{x} - \pdI{^2 u}{x^2} - \pdI{^4 u}{x^4} \label{e:modKS}  \\\nonumber
x \in [0,L], t \in[0,\infty) \\
u(0,t) = u(L,t) = 0 \nonumber\\
\left. \pdI{u}{x} \right|_{x=0} = \left. \pdI{u}{x} \right|_{x=L} = 0 \nonumber\\
u(x,0) = u_0(x) \nonumber
\end{gather}

Homogenenous Dirichlet and Neumann boundary conditions are used to make the system ergodic.  The parameter $c$ is added to demonstrate LSS. Sensitivity of the quantity of interests $\la \bar{u} \ra$ and $\la \bar{u^2} \ra$ to $c$ will be investigated, where:

\[
\la \bar{u} \ra \equiv \lim_{T \to \infty} \frac{1}{T} \int_0^T \bar{u} \ dt, \quad \bar{u} \equiv \frac{1}{L} \int_0^L u \ dx, \quad \bar{u^2} \equiv \frac{1}{L} \int_0^L u^2 \ dx
\]

\noindent We have chosen $L = 128$ to ensure the solution is chaotic \cite{Hyman:1986:KS}.  

The remainder of this paper is organized as follows: first, section \ref{s:numerics} discusses the numerical simulation used to compute solutions of the modified K-S equation.  Next, the reasons for the modifications to the K-S equation are discussed in more detail in section \ref{s:mod_KS}.  Thirdly, a brief summary of the LSS method will be given in section \ref{s:LSS}, followed by a presentation and a discussion of the gradients computed using LSS in section \ref{s:Results}.  Section \ref{s:conclusion} offers some concluding remarks and a discussion of future work.  

\section{Numerical Simulation}
\label{s:numerics}

The modified K-S equation was discretized with a 2nd order finite difference scheme.  We number the nodes $i = 0,1,2,...,n,n+1$, where $i = 0$ and $i = n+1$ are the boundary nodes and $i = 1,2,...,n$ denote the interior nodes.  Define $x_i = i\Delta x$, where $\Delta x = L/(n+1)$ is the spacing between each node, and $u_i = u(x_i)$.  The terms of the modified K-S equation were approximated as follows on the interior nodes:  

\begin{align*}
\left. \pdI{u}{x} \right|_i &\approx \frac{u_{i+1} - u_{i-1}}{2\Delta x}, \quad i=1,2,...,n \\
 \left. u\pdI{u}{x} \right|_i &=  \left. \frac{1}{2}\pdI{u^2}{x} \right|_{x_i} \approx \frac{u_{i+1}^2 - u_{i-1}^2}{4\Delta x}, \quad i=1,2,...,n \\
 \left. \pdI{^2 u}{x^2} \right|_i &\approx \frac{u_{i+1} - 2u_i + u_{i-1}}{\Delta x^2}, \quad i=1,2,...,n \\
 \left. \pdI{^4 u}{x^4} \right|_i &\approx \frac{u_{i-2} -4u_{i-1} +6u_i -4u_{i+1} + u_{i+2} }{\Delta x^4}, \quad i=2,3,...,n-1
\end{align*}

To enforce the homogeneous Dirichlet boundary conditions we set $u_0 = u_{n+1} = 0$.  To enforce the homogeneous Neumann boundary conditions, we make use of ghost nodes, setting $u_{-1} = u_1$ and $u_{n+2} = u_n$.  This ensures that the central difference approximation of $\pdI{u}{x}$ is zero at nodes $0$ and $n$, which correspond to $x=0$ and $x=L$, respectively.  Therefore:

\begin{align*}
\left. \pdI{u}{x} \right|_1 &\approx \frac{u_2 }{2\Delta x}, \quad \left. \pdI{u}{x} \right|_n \approx -\frac{u_{n-1}}{2\Delta x} \\
 \left. u\pdI{u}{x} \right|_1 &\approx \frac{u_2^2}{4\Delta x}, \quad  \left. u\pdI{u}{x} \right|_n \approx -\frac{u_{n-1}^2}{4\Delta x} \\
\left. \pdI{^2 u}{x^2} \right|_1 &\approx \frac{u_2 - 2u_1}{\Delta x^2}, \quad 
\left. \pdI{^2 u}{x^2} \right|_n \approx \frac{u_{n-1} - 2u_n}{\Delta x^2} \\
 \left. \pdI{^4 u}{x^4} \right|_1 &\approx \frac{7u_1 -4u_2 + u_3 }{\Delta x^4}, \quad
  \left. \pdI{^4 u}{x^4} \right|_n \approx \frac{7u_n -4u_{n-1} + u_{n-2} }{\Delta x^4}\\
 \left. \pdI{^4 u}{x^4} \right|_2 &\approx \frac{-4u_1 + 6u_{2} -4u_{3} + u_{4} }{\Delta x^4}, \quad
  \left. \pdI{^4 u}{x^4} \right|_{n-1} \approx \frac{-4u_n + 6u_{n-1} -4u_{n-2} + u_{n-3} }{\Delta x^4}
\end{align*}

\begin{figure} 
\centering
\subfloat[$J=\bar{u}$]
{\includegraphics[width=3.2in]{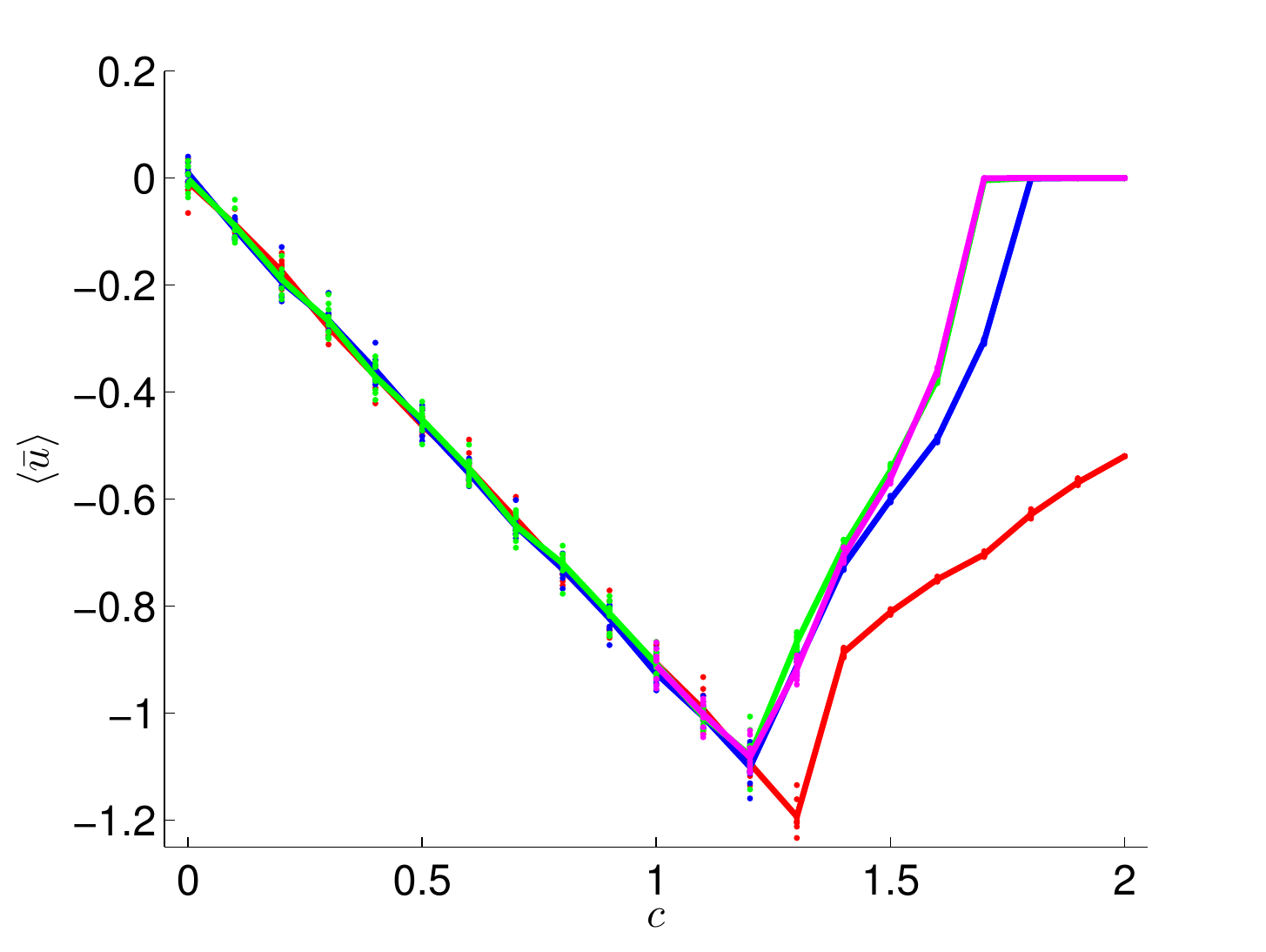}}\\
\subfloat[$J=\bar{u^2}$]
{\includegraphics[width=3.2in]{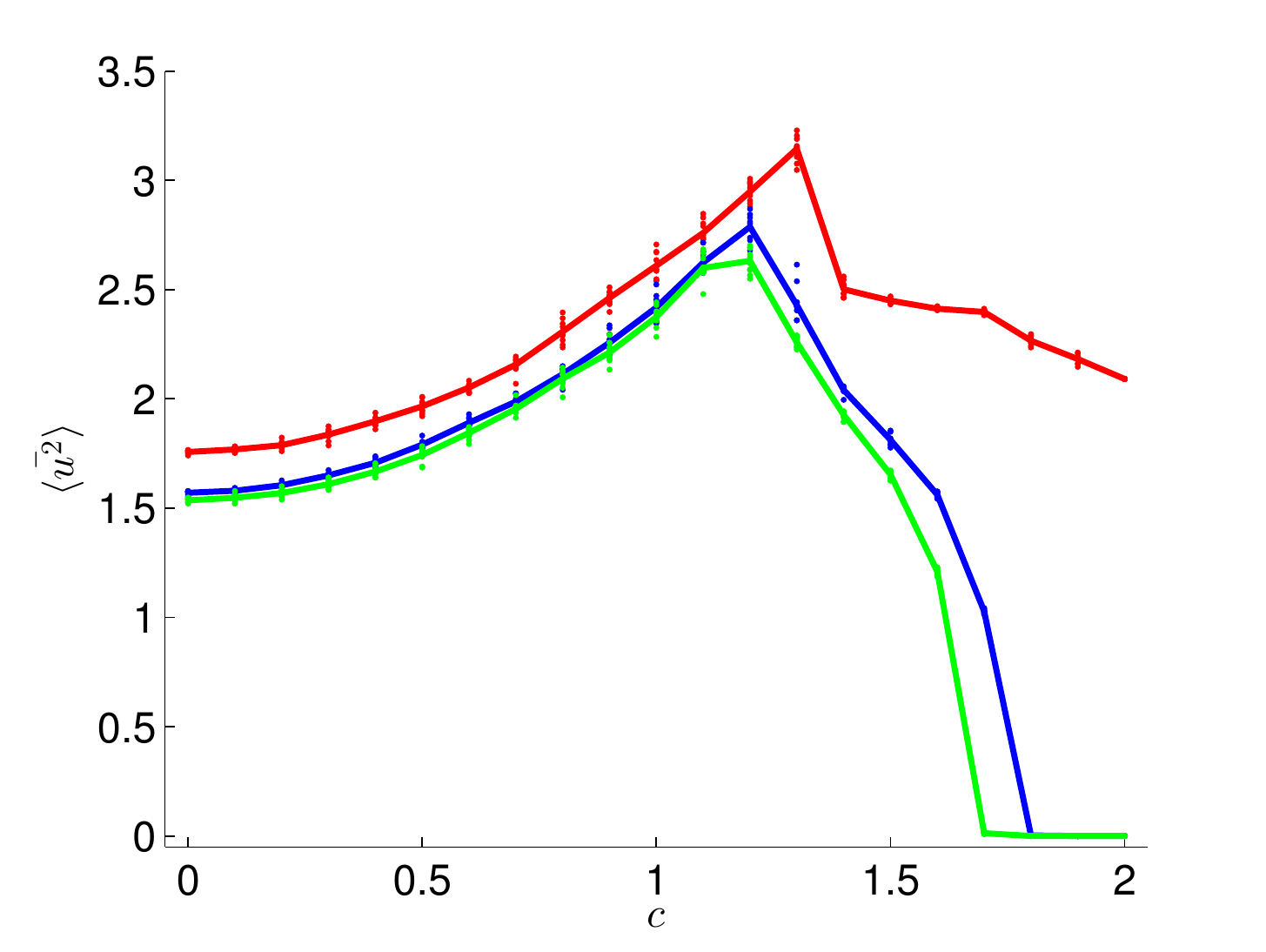}}

\caption{Two different objective functions $\la J \ra$ versus $c$.  Dots represent each realization, lines represent the average of 10 realizations.  Different colors correspond to different spatial discretizations, with red for $n = 127$, blue for $n = 255$, green for $n = 511$, and maroon for $n = 700$, where $n$ is the number of nodes.  Each realization was run for $1000$ time units before averaging was begun.  Averages were taken over $2000$ time unit intervals.  } 
\label{f:obj_vs_c}
\end{figure}

The coarsest mesh we used to solve the modified K-S equation is a uniform mesh with 127 interior nodes, as  in Brummitt and Sprott \cite{Brummitt:2009:simplePDE}. From figure \ref{f:obj_vs_c}, we can see that while this mesh is accurate for $0 \le c \le 1.2$, where the mean values of $\la \bar{u} \ra$ never vary more than 5\%.  However. The approximation is poor for larger magnitude values of $c$ and for computing $\la \bar{u^2} \ra$.  A mesh with $n=511$ interior nodes and $\Delta x = 0.25$ is used for these computations, as this grid resolution appears to offer sufficient accuracy in figure \ref{f:obj_vs_c}.

Time integration was mainly conducted with a 3rd order accurate implicit-explicit Runge-Kutta time integration scheme, IMEXRK34S[2R]L$\alpha$ \cite{Cava:2013:IMEX}, with the Butcher tableau shown in table \ref{t:imex}. The 2nd and 4th order derivatives in equation \eqref{e:modKS} were integrated with the Diagonally Implicit Runge-Kutta (DIRK) scheme, and the convective terms,  $-(u+c)\pdI{u}{x}$, were integrated with the Explicit Runge-Kutta (ERK) scheme.  For the results presented in this paper, a time step of $\Delta t =0.1$ was used unless otherwise stated.  

\begin{table}
\centering
\begin{tabular}{c|cccc}
0 & 0 &  &  &  \\
1/3 & 0 & 1/3 &  &  \\
1 & 0 & 1/2 & 1/2 &  \\
1 & 0 & 3/4 & -1/4 & 1/2 \\\hline
  & 0 & 3/4 & -1/4 & 1/2 \\
\end{tabular}
\hspace{0.5in}
\begin{tabular}{c|cccc}
 0 & 0 & & & \\
 1/3 & 1/3 & 0 & & \\
 1 & 0 & 1 & 0 & \\
 1 & 0 & 3/4 & 1/4 & 0 \\\hline
   & 0 & 3/4 & -1/4 & 1/2 \\
\end{tabular}
\caption{Butcher Tableau for the Diagonally Implicit Runge-Kutta (DIRK) scheme (left) and the Explicit Runge-Kutta (ERK) scheme (right) used by IMEXRK34S[2R]L$\alpha$.  The scheme was designed to maximize the accuracy of the explicit portion of the solver \cite{Cava:2013:IMEX}.  In the case of this paper, the explicit portion is the convective terms of \eqref{e:modKS}, $-(u+c)\pdI{u}{x}$.  }
\label{t:imex}
\end{table}

Some of the ensemble average computations (see section \ref{s:mod_KS}) were conducted using MATLAB's ODE45 hybrid 4th/5th order Runge-Kutta scheme and a time step size of $\Delta t = 0.2$ to ensure stability.  Solutions to the modified K-S equation used for the work presented in this paper range from 100 to 4000 time units.  

Finally, unless otherwise stated, the initial condition $u_0(x)$ was formed by randomly selecting numbers at each spatial node from a uniform distribution between $u = -0.5$ and $u = 0.5$.  

\section{The Modified Kuramoto-Sivashinsky Equation}
\label{s:mod_KS}

\subsection{Boundary Conditions}

For LSS to compute accurate gradients, the system being analyzed must be ergodic; the long time behavior of the system is independent of initial conditions. The K-S equation with periodic boundary conditions is not ergodic, and to show this we consider the spatial average of equation \eqref{e:modKS}:

\[
\frac{1}{L}  \int_0^L \left( \pdI{u}{t} \right) dx =  \frac{1}{L}  \int_0^L \left( -(u + c) \pdI{u}{x} - \pdI{^2 u}{x^2} - \pdI{^4 u}{x^4} \right) dx 
\]

Taking the time derivative outside of the integral, using the notation $\bar{u} = \frac{1}{L}\int_0^L u \ dx$, and multiplying both sides of the equation by $L$, we obtain:

\begin{equation}
L\pdI{\bar{u}}{t}  = -\frac{1}{2} \left. u^2 \right|_0^L - c \left. u \right|_0^L -\left. \pdI{u}{x} \right|_0^L -\left. \pdI{^3 u}{x^3} \right|_0^L
\label{e:KSint}
\end{equation}

For periodic boundary conditions, $u$ and all of its derivatives are equal at $x=0$ and $x=L$ for all time.  Therefore equation \eqref{e:KSint} becomes:

\[
\pdI{\bar{u}}{t}  = 0
\]

\begin{figure} \centering
\subfloat[$T_0 = 500$]
{\includegraphics[width=3.2in]{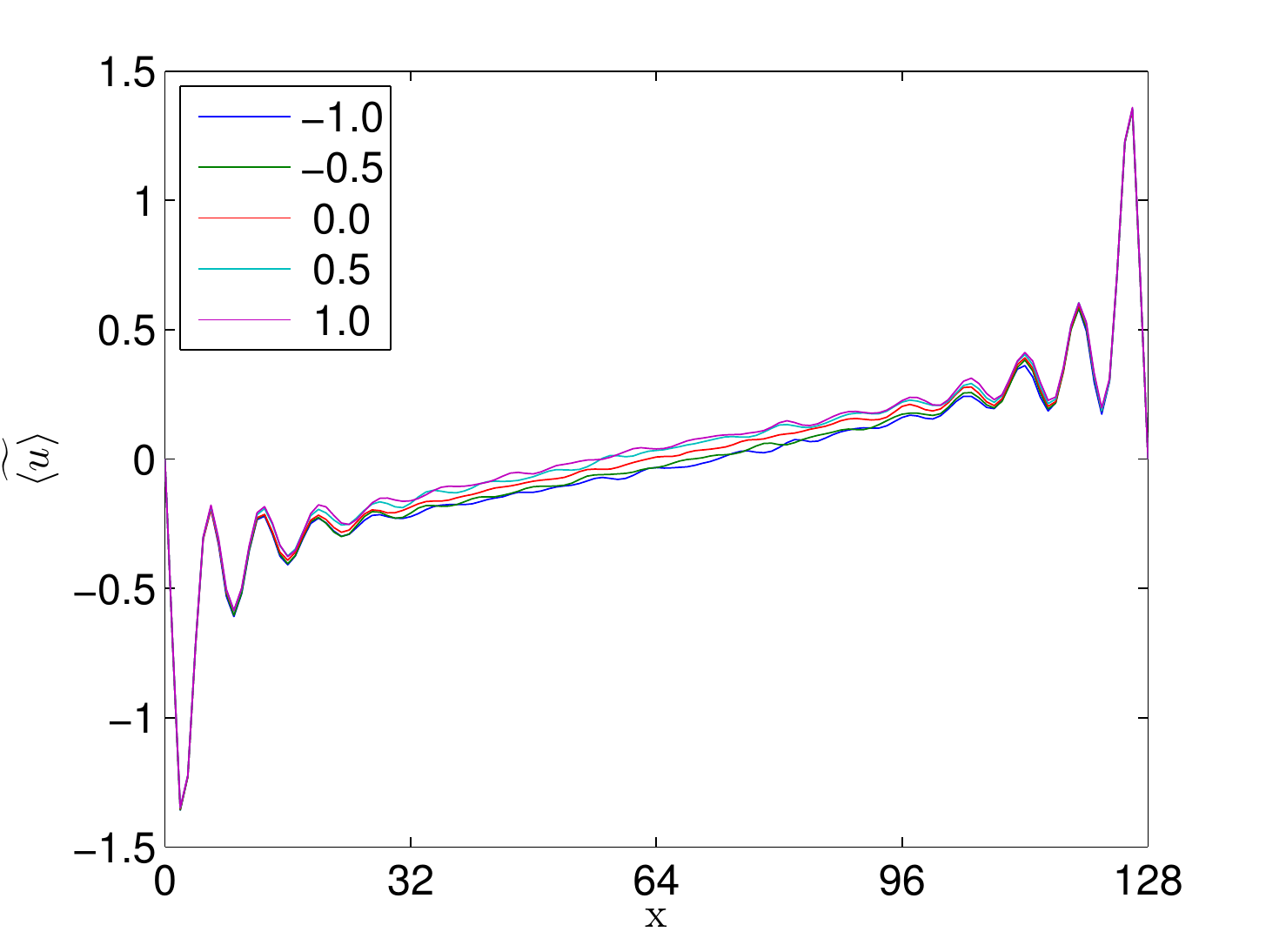}}\\
\subfloat[$T_0 = 1000$]
{\includegraphics[width=3.2in]{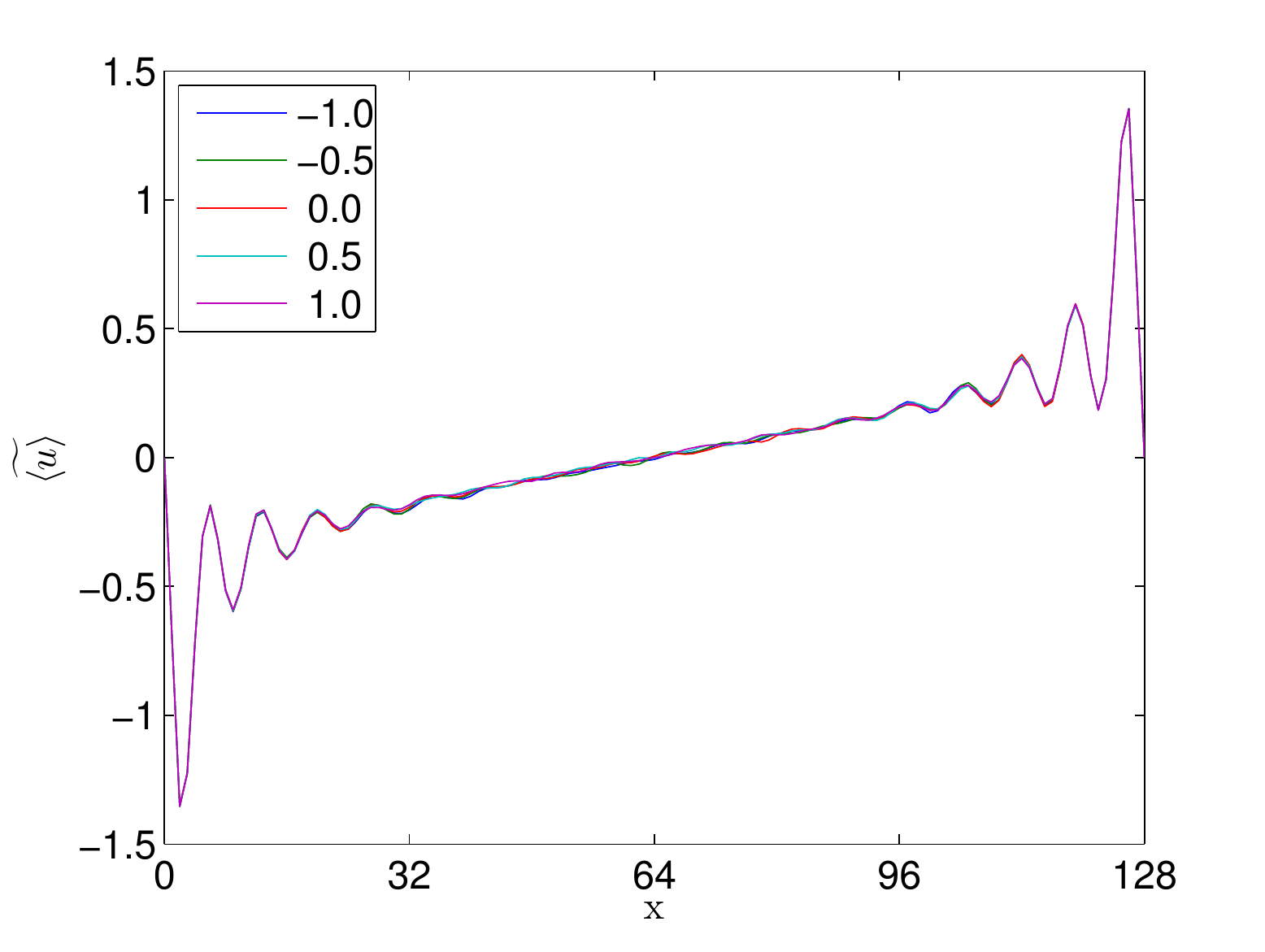}}
\caption{The ensemble averaged time averaged solution, $\widetilde{\la u \ra}$, for $c = 0$ for initial conditions with five different means $\bar{u}_0$. Both solutions were run for $T_0$ time units before the averaging was started.  The averaging interval was 100 time units long and 4000 realizations were used for both ensemble averages. All five initial conditions were formed by randomly selecting numbers at each grid point from a uniform distribution between $u = -0.5$ and $u = 0.5$ and then changing the mean.  }
\label{f:ergodicity}
\end{figure}

\noindent This means that $\bar{u}$ is always equal to the average of the initial condition $u_0(x)$. Therefore, our choice of the initial condition dictates the long time behavior of our quantity of interest $\bar{u}$ and the system is not ergodic.  On the other hand if homogeneous Dirichlet and Neumann boundary conditions are used, ergodic behavior can shown from numerical solutions.  Ergodicity can be observed in the behavior of the ensemble averaged time averaged solution, defined as
\[
\widetilde{\la u \ra} \equiv \frac{1}{K} \sum_{k=1}^K \la u \ra
\]
\noindent where $\la u \ra$ is the time average of $u(x,t)$.  The ensemble average is conducted over $K$ realizations, each starting from a different initial condition, comprised of the random noise described in section \ref{s:numerics} shifted to some non-zero mean.  Figure \ref{f:ergodicity} shows that $\widetilde{\la u \ra}$ converges to a single function for initial conditions with different means.  The initial condition also contained random noise like that described in section \ref{s:numerics}. The random noise component of the initial condition is comprised of many different functions, which means that the convergence shown in figure \ref{f:ergodicity} strongly implies that the system is ergodic.  Note that similar convergence of the solutions was observed at a number of different values of $c$, suggesting that the modified K-S equation is ergodic over a wide range of values of $c$.  

Finally, note that the initial condition $u_0(x) = 0$ results in a trivial solution $u(x,t)=0$.  However, this trivial solution is unstable; any small perturbation to it will lead to a non-trivial, chaotic solution \cite{Hyman:1986:KS}.  One can think of the trivial solution to the modified K-S equation as an unstable fixed point, like $(x,y,z) = (0,0,0)$ in the Lorenz System.  

\subsection{Additional Linear Convection Parameter $c$}

\begin{figure} 
\centering
\subfloat[$c<1.3$]
{\includegraphics[width=3.2in]{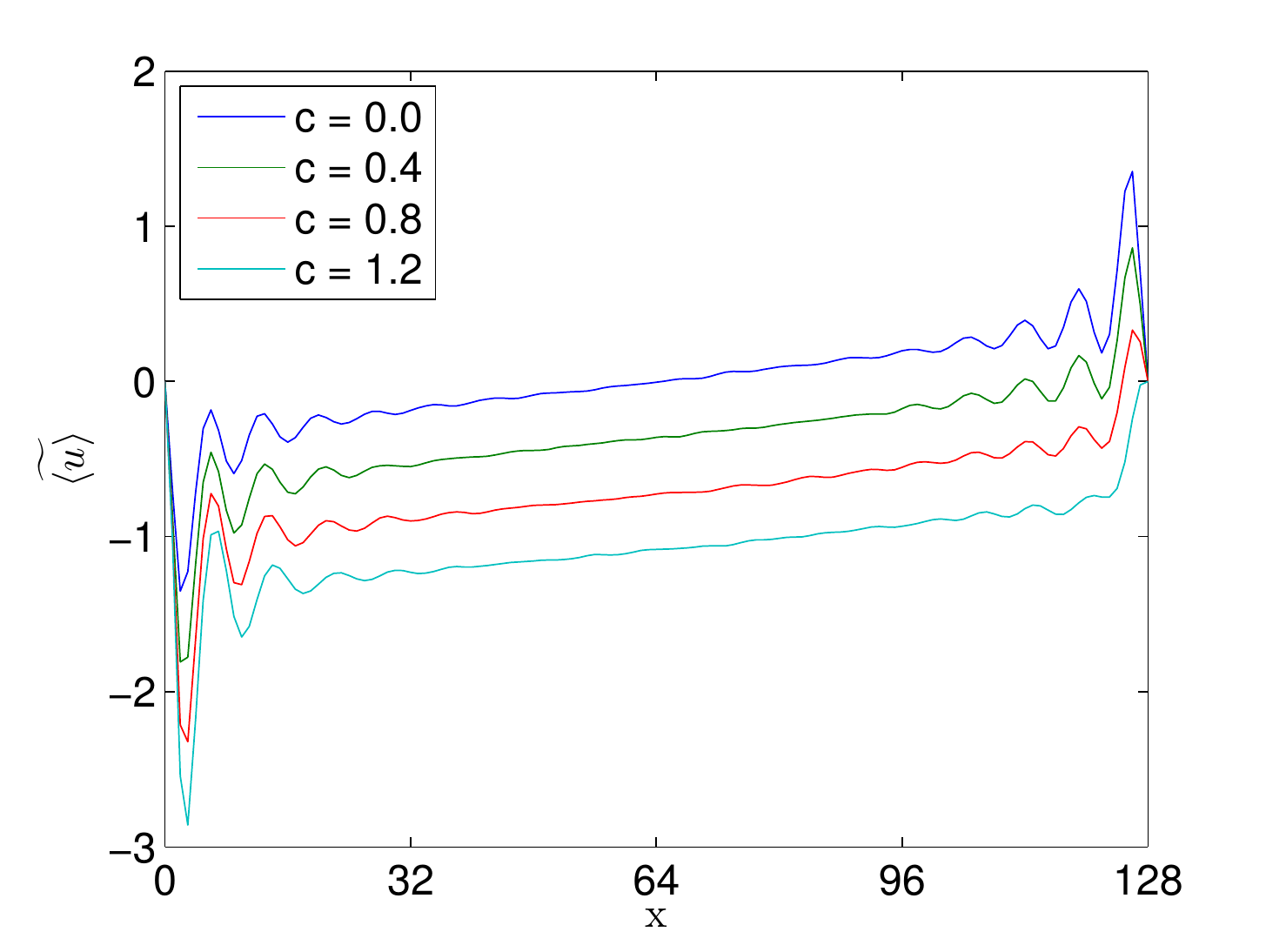}}\\
\subfloat[$c>1.3$]
{\includegraphics[width=3.2in]{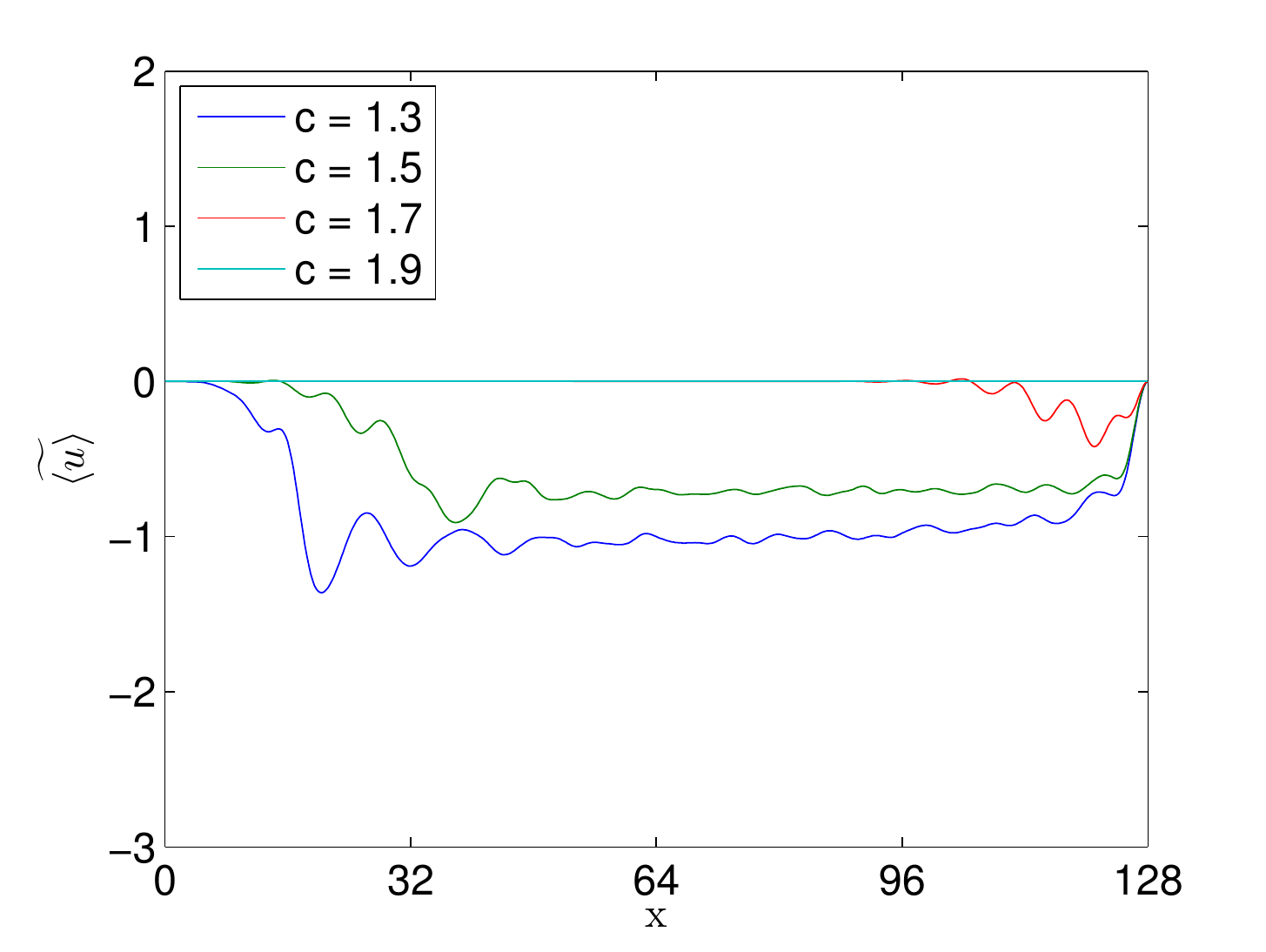}}
\caption{The ensemble averaged, time averaged solution,
 $\widetilde{\la u \ra}$ for different values of $c$. The averaging interval was 100 time units long and 4000 realizations were used for both ensemble averages. }
\label{f:ensemble_c}
\end{figure}

To see the effect of the parameter $c$, we again consider ensemble averaged time averaged solutions $\widetilde{\la u \ra}$. Figure \ref{f:ensemble_c} shows that as $c$ is increased, $\widetilde{\la u \ra}$ is decreased at all values of $x$ other than the boundaries, where the Dirichlet boundary conditions are imposed.  

This decrease in $\widetilde{\la u \ra}$ occurs because increasing $c$ leads to increased linear convection in the positive $x$ direction, transporting negative $u$ from one side of the domain to the other and transporting positive $u$ out of the domain.  

This trend can be seen in figure \ref{f:obj_vs_c}, which shows how time and space averaged $u$, $\la \bar{u} \ra$, varies with $c$. From $c=0$ to $c\approx 1.2$, we see the behavior shown in figure \ref{f:ensemble_c} and $\la \bar{u} \ra$ decreases linearly with $c$.  Note that because the the boundary conditions are symmetric, $\la \bar{u} \ra$ as a function of $c$ is anti-symmetric.  Decreasing $c$ from $c=0$ results in an linear increase in $\la \bar{u} \ra$, until $c\approx -1.2$.

\begin{figure} 
\centering
\includegraphics[scale=0.4]{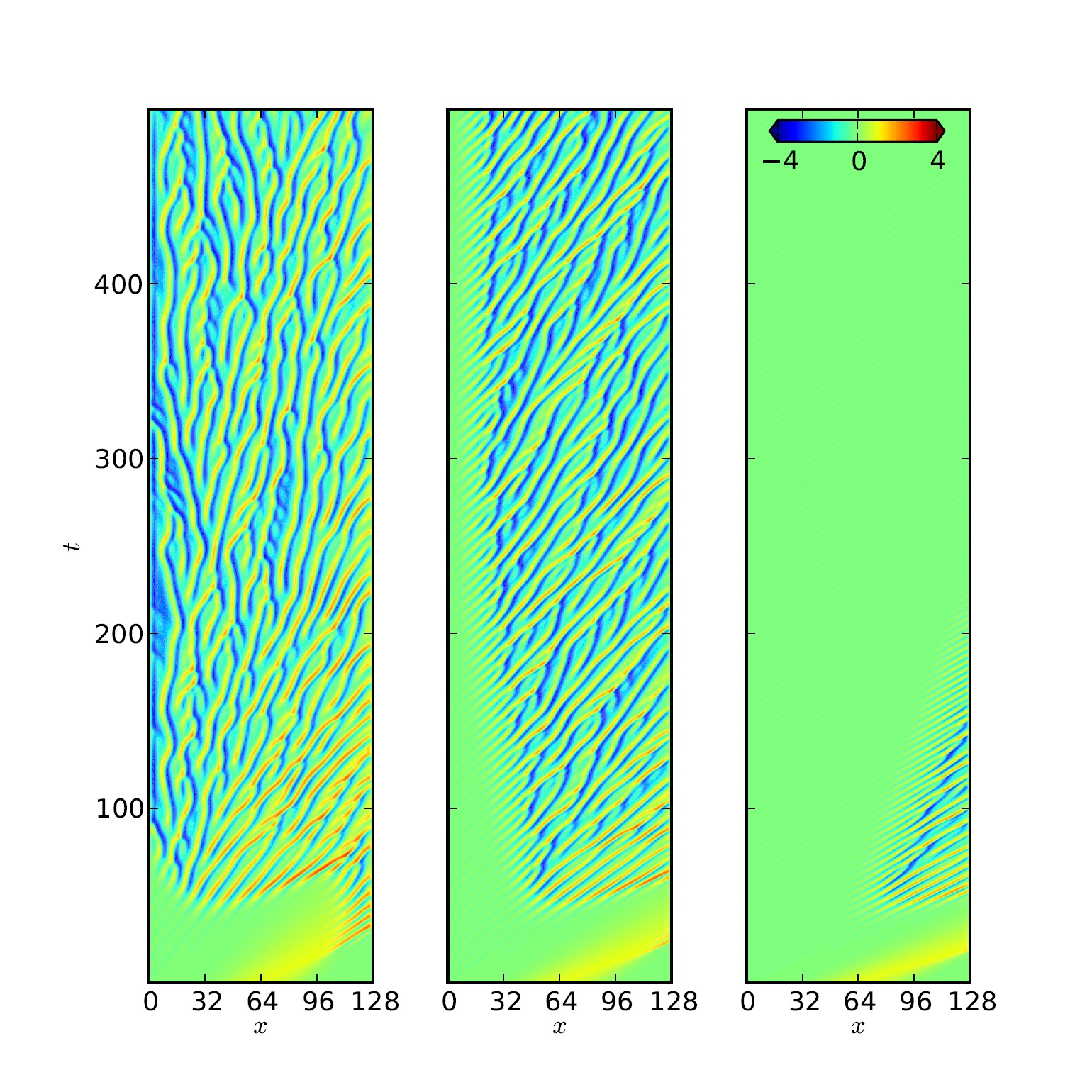}
\caption{From left to right: $x$ vs. $t$ plots for $u(x,t)$ for $c = 0.8$ (light turbulence dominated regime), $c = 1.4$ (convection dominated regime), and $c=2.0$ (steady regime).  All three histories were computed with $\Delta x = 0.25$.  The initial condition for all three histories was $u_0(x) = e^{\frac{-(x-64)^2}{512}}$. } 
\label{f:histories}
\end{figure}

For $c>1.2$ (or $c<-1.2$) we see a change in the slope of $\la \bar{u} \ra$.  Another change occurs around $c=1.8$.  These changes in the trend of $\la \bar{u} \ra$ are a result of changes in the dynamics of the modified K-S equation.  Example of solutions $u(x,t)$ in these regimes, $0\le c \le 1.2$, $1.2\le c \le 1.8$, and $c > 1.8$ are shown in figure \ref{f:histories}.  

In the first regime, $0\le c \le 1.2$, the ``light turbulence dominated regime'', we see the chaotic spatio-temporal structures referred to in many past studies of the K-S.  In the left-most plot of figure \ref{f:histories} we see that some of the structures between roughly $x = 64$ and $x = 128$ are convecting towards $x=128$, due to the linear convection term.  

As $c$ is increased, convection in the positive $x$-direction is increased.  This can be seen visually from the tilt or slope of the spatio-temporal structures in all three $x-t$ diagrams in figure \ref{f:histories}.  The smaller the slope of the structure, the faster the structure is being convected. 

In the second regime, $1.2\le c \le 1.8$, the ``convection dominated regime'', All spatio-temporal structures are convecting towards $x=128$, as seen in the center plot of figure \ref{f:histories}.  Additionally, a region of $u=0$ now exists near $x=0$.  As $x$ increases, the spatio-temporal structures grow, until $x\approx 30$ for $c=1.4$ (figure \ref{f:histories}).  The size of this region grows as $c$ is increased.  

In the final regime, $c > 1.8$, the ``steady regime'', the chaotic structures are convected out of the domain, as in the right-most plot in figure \ref{f:histories}.  As $c$ is increased, the structures are convected out of the spatial domain faster.  The trivial solution has changed from an unstable fixed point to a stable fixed point, and the solution becomes $u(x,t)=0$ after some time.  This is different from solutions in the first two regimes, both of which are on chaotic attractors, as indicated by ergodicity and presence of spatio-temporal chaos.

\section{The Least Squares Shadowing Method}
\label{s:LSS}

\noindent Say we are interested in the sensitivity of our long-time averaged quantity $\la \bar{u} \ra$ to the parameter $c$:

\[
\ddI{\la \bar{u} \ra}{c} = \ddI{}{c} \left( \lim_{T \to \infty} \frac{1}{T} \int_0^T \bar{u} \ dt \right) 
\]

\noindent For non-chaotic solutions of the K-S equation solved from the {\bf initial value problem}, we can exchange the derivative and the time limit:

\begin{equation}
\ddI{\la \bar{u} \ra}{c} =\lim_{T \to \infty} \frac{1}{T} \int_0^T \pdI{\bar{u}}{u} \pdI{u}{c} \ dt
\label{e:sens}
\end{equation}

\noindent Where 

\begin{equation}
\pdI{u}{c} = \lim_{\varepsilon \to 0} \frac{u(t;c+ \varepsilon) - u(t;c)}{\varepsilon}
\label{e:deriv}
\end{equation}

We can solve for the {\it tangent solution}, $v \equiv \pdI{u}{c}$, with the linearization of equation \eqref{e:modKS}, also called the {\it tangent equation}.  However, if $u(t;c)$ is chaotic, then equations \eqref{e:sens} and \eqref{e:deriv} do not hold and the tangent solution, $v$. diverges exponentially as the simulation time $T$ is increased \cite{Lea:2000:climate_sens,Wang:2013:LSS1}.  

\begin{figure}
\centering
\includegraphics[width=0.6\textwidth]{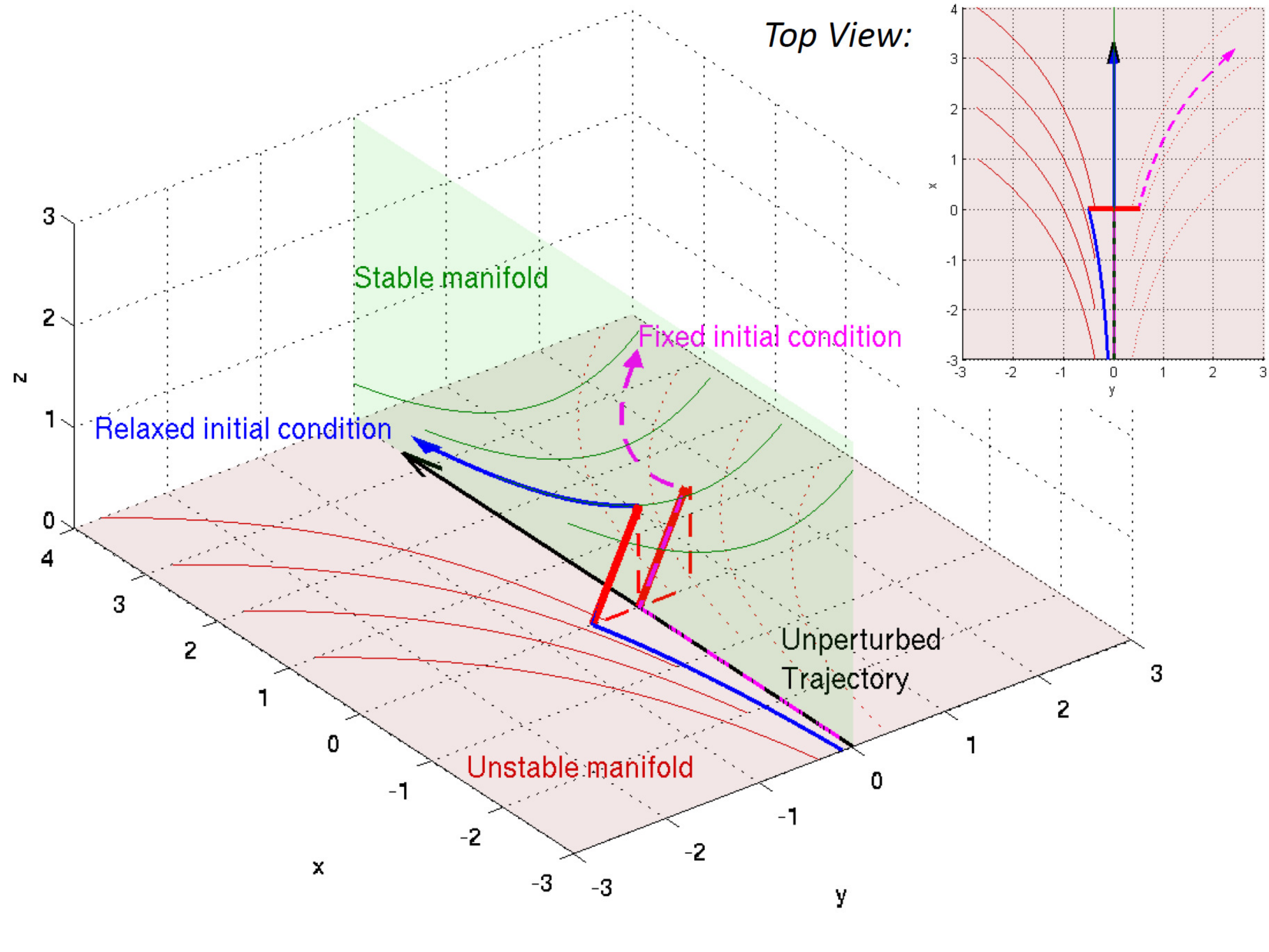}
\caption{Phase space trajectory of a chaotic dynamical system.  The unstable manifold, in red, is the space of all Lyapunov covariant vectors corresponding to positive exponents. The stable manifold, in green, corresponds to the space of all covariant vectors associated with negative exponents.  A perturbation to the system (in red) has components in both manifolds, and the unstable component causes the perturbed trajectory (pink) to diverge exponentially from the unperturbed trajectory (in black).  LSS chooses a perturbed trajectory with a different initial condition (in blue) that does not diverge from the unperturbed trajectory.  }
\label{f:LSS_schematic}
\end{figure}

For a steady or periodic solution, a slight change $\varepsilon$ in $c$ results in a slightly different solution $u(t;c+ \varepsilon)$ for all time and the tangent solution, $v$, can be computed accurately.  However, the positive Lyapunov exponent(s) present on strange attractors ensure that the solutions $u(t;c+ \varepsilon)$ and $u(t;c)$ will be very different after some time, as illustrated in figure \ref{f:LSS_schematic}.  We see that if the perturbed solution, or phase space trajectory, has the same initial condition as the unperturbed trajectory, the two trajectories diverge exponentially.  This exponential divergence of the two trajectories causes the tangent solution, $v$, to diverge, resulting in the issues with traditional sensitivity analysis mentioned in the introduction.

However, the assumption of ergodicity means that it is not necessary to compare a perturbed and an unperturbed trajectory with the same initial condition if the quantities of interest are statistics of the system such as long time averages.  Therefore, an initial condition can be chosen such that the perturbed and unperturbed trajectories do not diverge, resulting in the blue trajectory in figure \ref{f:LSS_schematic}. The existence of this trajectory, called a ``shadow trajectory'', follows from the shadowing lemma \cite{Pilyugin:1999:shadow}: 

\begin{quote}
{\it
For any $\delta > 0$ there exists $\varepsilon > 0$, such that for every ``$\varepsilon$-pseudo-solution'' $u_{\varepsilon}$ that satisfies $\|d u_{\varepsilon} / d t - \mathcal{R}(u_{\varepsilon})\|<\varepsilon$, $0 \le t \le T$, there exists a true solution $u$ and a time transformation $\tau(t)$, such that $\|u(\tau(t))-u_{\varepsilon}(t)\| < \delta$, $|1-d\tau/dt|<\delta$ and $d u/d\tau - \mathcal{R}(u) = 0$, $0 \le \tau \le \mathcal{T}$. 
}
\end{quote}

Where the norm $\|\cdot\|$ refers to distance in phase space. For the K-S equation, the operator $\mathcal{R}$ is comprised of the spatial derivative operators on the right hand side of equation \eqref{e:modKS}.  

\begin{figure}
\centering
\includegraphics[width=0.35\textwidth]{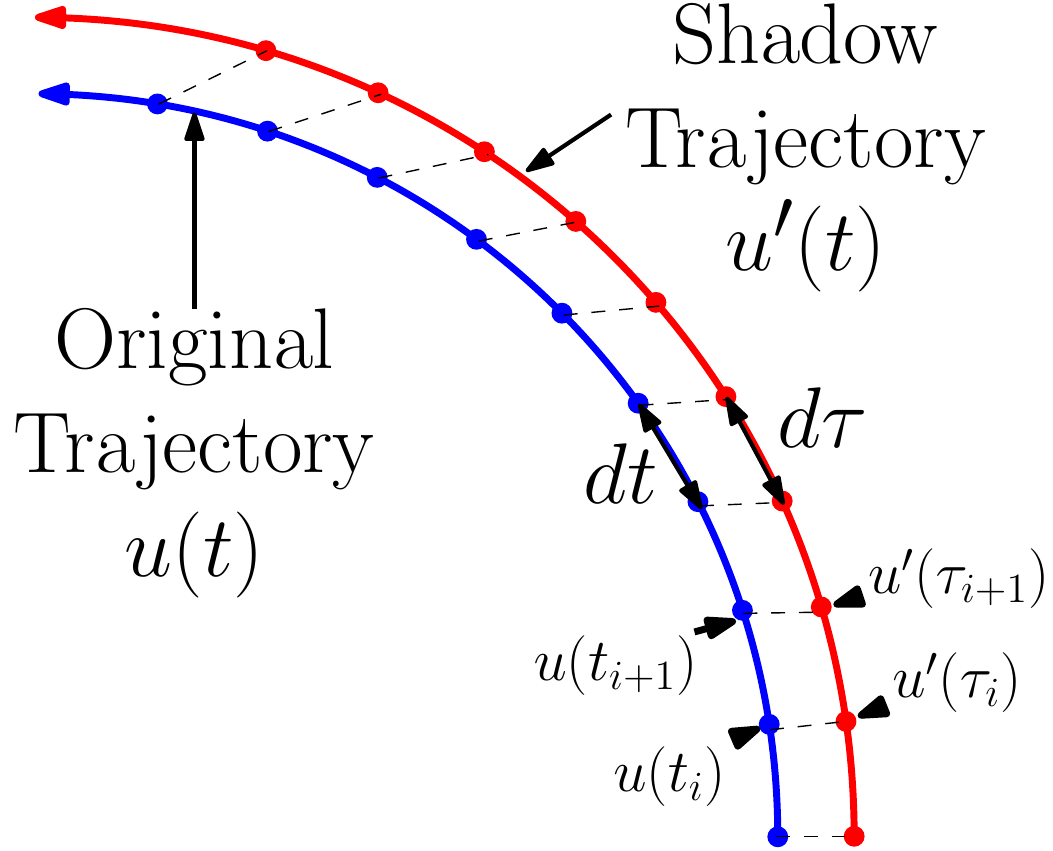}\\
\includegraphics[width=0.35\textwidth]{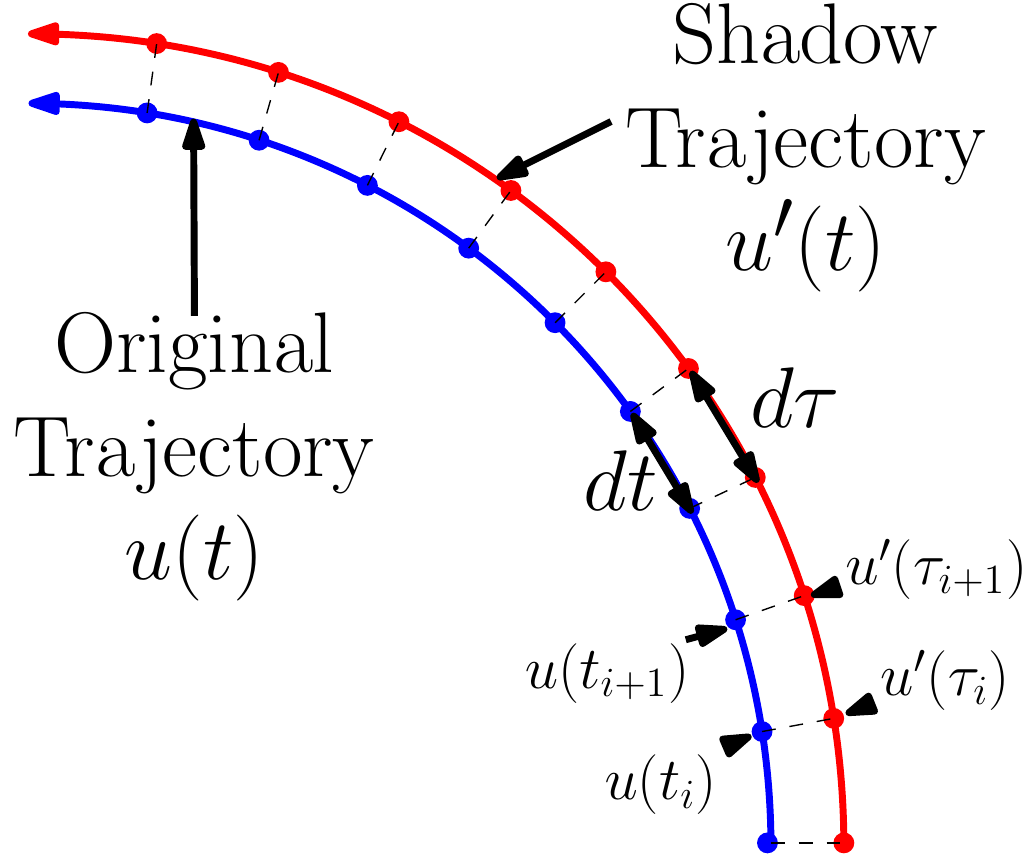}
\caption{TOP: Original and shadow phase space trajectories without any time transformation ($d\tau/dt = 1$).  BOTTOM: Original and shadow phase space trajectories with a time transformation $d\tau/dt = 1 + \eta$ that minimizes the distance between the two trajectories in phase space for all time.  }
\label{f:time_dilation}
\end{figure}

The time transformation alluded to in the shadowing lemma is required to deal with the zero (neutrally stable) Lyapunov exponent on the strange attractor.  The need for this transformation is clarified in figure \ref{f:time_dilation}.  The time transformation, referred to as ``time dilation'' in this paper and other LSS literature, is required to keep a phase space trajectory and its shadow trajectory close (in phase space) for infinite time.

If we use two solutions that shadow one another in equation \eqref{e:deriv}, \ref{e:sens} can be used to compute accurate sensitivities \cite{Wang:2013:LSSthm}.  $v \equiv \pdI{u}{c}$, called the ``shadowing direction'' can be computed by solving the following optimization problem with the tangent equation as a constraint:

\begin{gather}
 \min_{v,\eta} \frac{1}{2T}\int_0^T v^2 + \alpha^2 \eta^2 dt, \nonumber\\ \text{s.t.} \quad \frac{dv}{dt} = \frac{\partial \mathcal{R}}{\partial u} v + \frac{\partial \mathcal{R}}{\partial c} + \eta \mathcal{R}(u;c) \label{e:opt_problem}\\ \quad 0<t<T \nonumber
\end{gather}

\noindent where $\eta = d\tau/dt - 1$ is the time dilation term, which corresponds to the time transformation from the shadowing lemma and $\alpha^2$ is a weighting parameter for the optimization problem.  More details on the implementation of LSS can be found in \cite{Wang:2013:LSS2}.  

\section{Results}
\label{s:Results}

\subsection{Shadowing for the Modified Kuramoto-Sivashinsky Equation}

\begin{figure}
\centering
\includegraphics[scale=0.2]{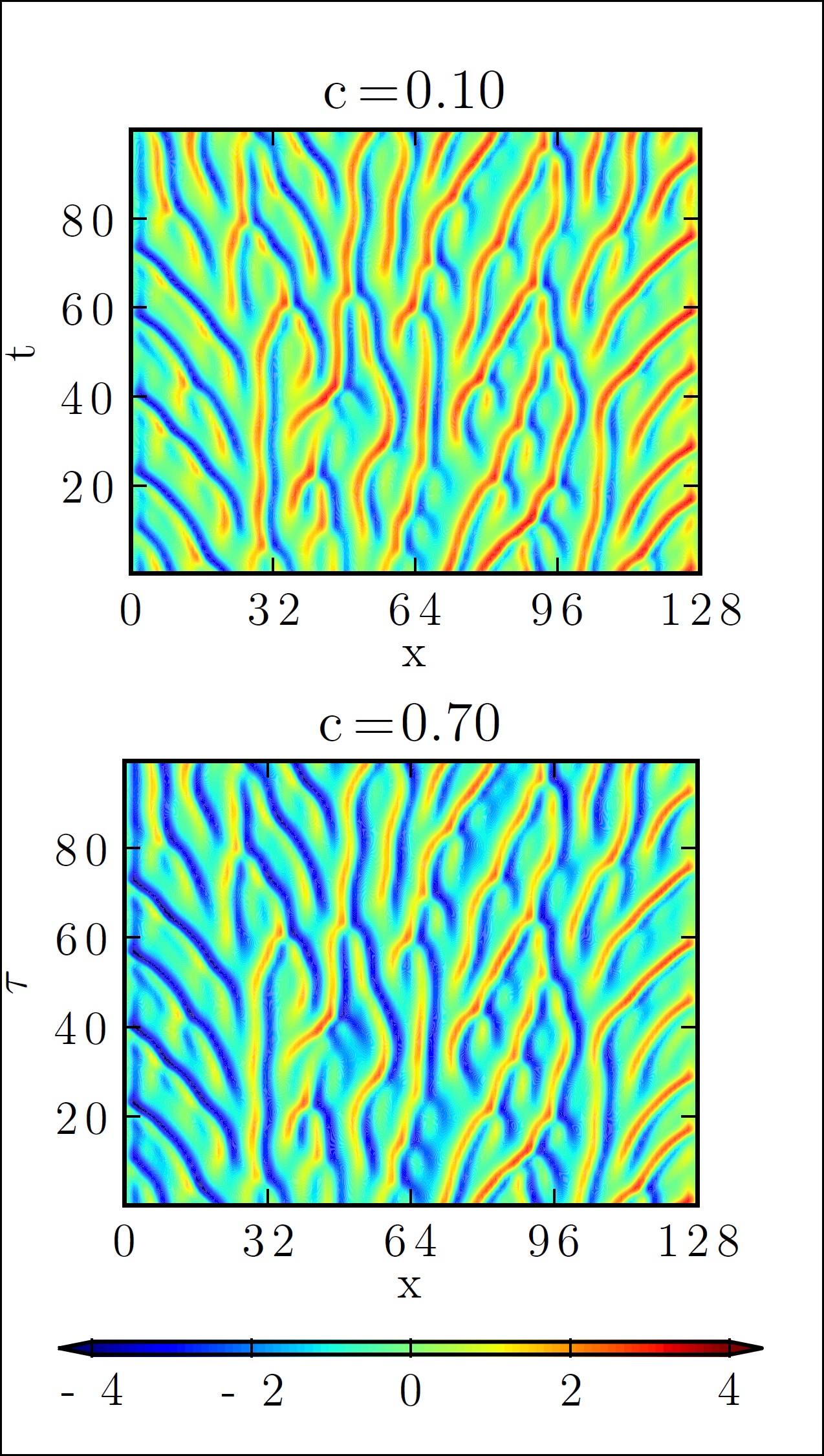}
\caption{TOP: $x$ vs. $t$ plot for $u(x,t)$ for $c = 0.1$.  BOTTOM: $x$ vs. $\tau$ plot for the corresponding shadow trajectory $u_s(x,\tau)$ with $c = 0.7$.  This shadow trajectory was computed using the algorithm outlined in \cite{Blonigan:2013:LSS}.}  
\label{f:shadow}
\end{figure}

The concept of a shadow trajectory in phase space is easy to visualize for low dimensional dynamical systems such as the three degree of freedom Lorenz system.  This is not the case for high dimensional systems, like our simulation of the modified K-S equation, which has between 127 and 511 dimensions.  To understand what shadowing entails for a high dimensional ODE or a PDE, consider figure \ref{f:shadow}.  We see that the shadow trajectory $u_s(x,\tau)$ has a lower average value than $u(x,t)$ from the coloring of the contours.  Additionally, we see that $u_s(x,\tau)$ has spatio-temporal structures that has very similar to $u(x,t)$.  This is because the spatio-temporal structures of $u_s(x,\tau)$ shadow those in $u(x,t)$.  Also, note that the time scale of $u_s(x,\tau)$ is indistinguishable from $u(x,t)$, indicating that there is very little time dilation.  

\subsection{Least Squares Shadowing Sensitivity Analysis}

\begin{figure} 
\centering
\subfloat[$J=\bar{u}$]{
\includegraphics[width=3.2in]{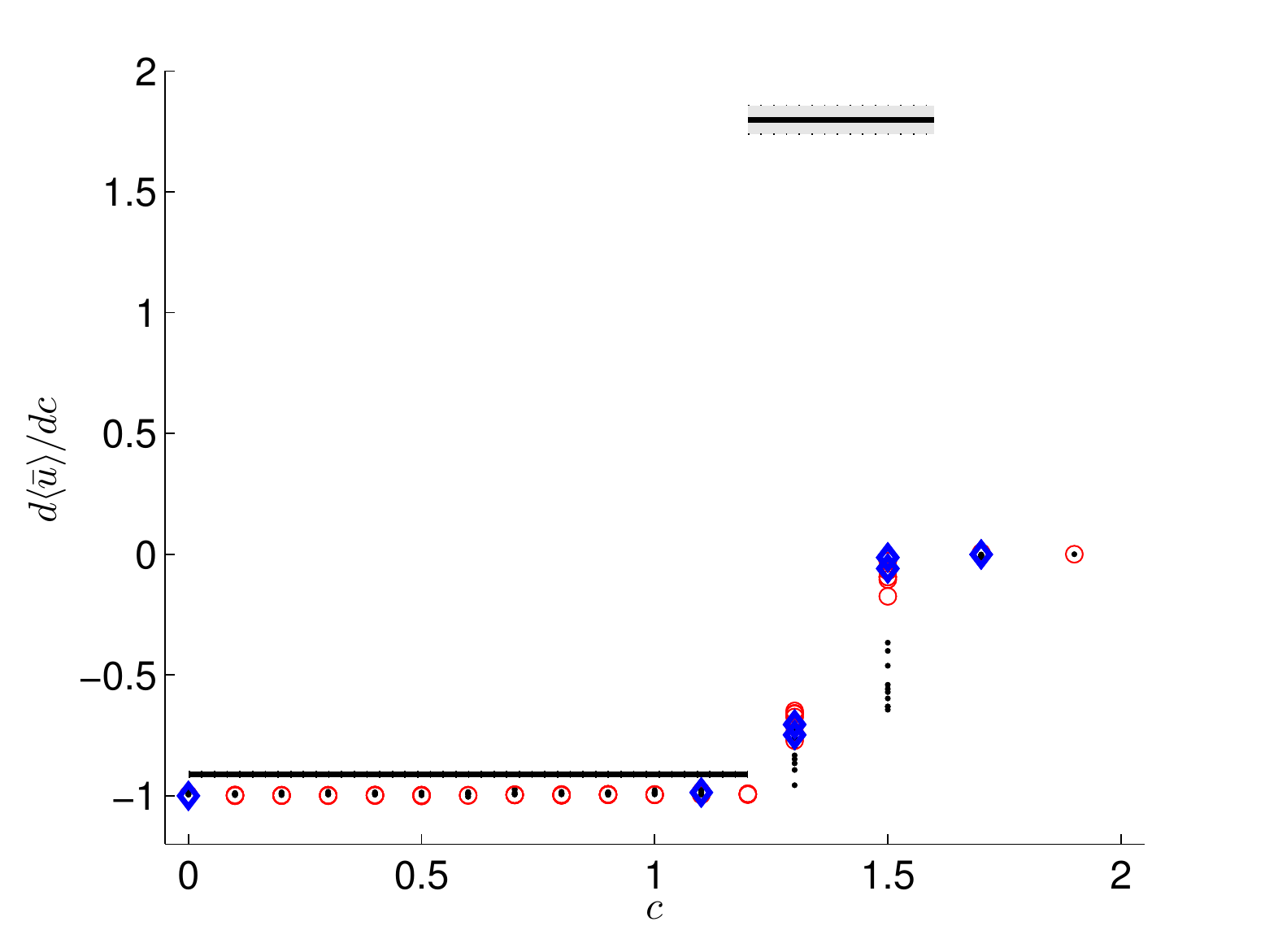}}\\
\subfloat[$J=\bar{u^2}$]{
\includegraphics[width=3.2in]{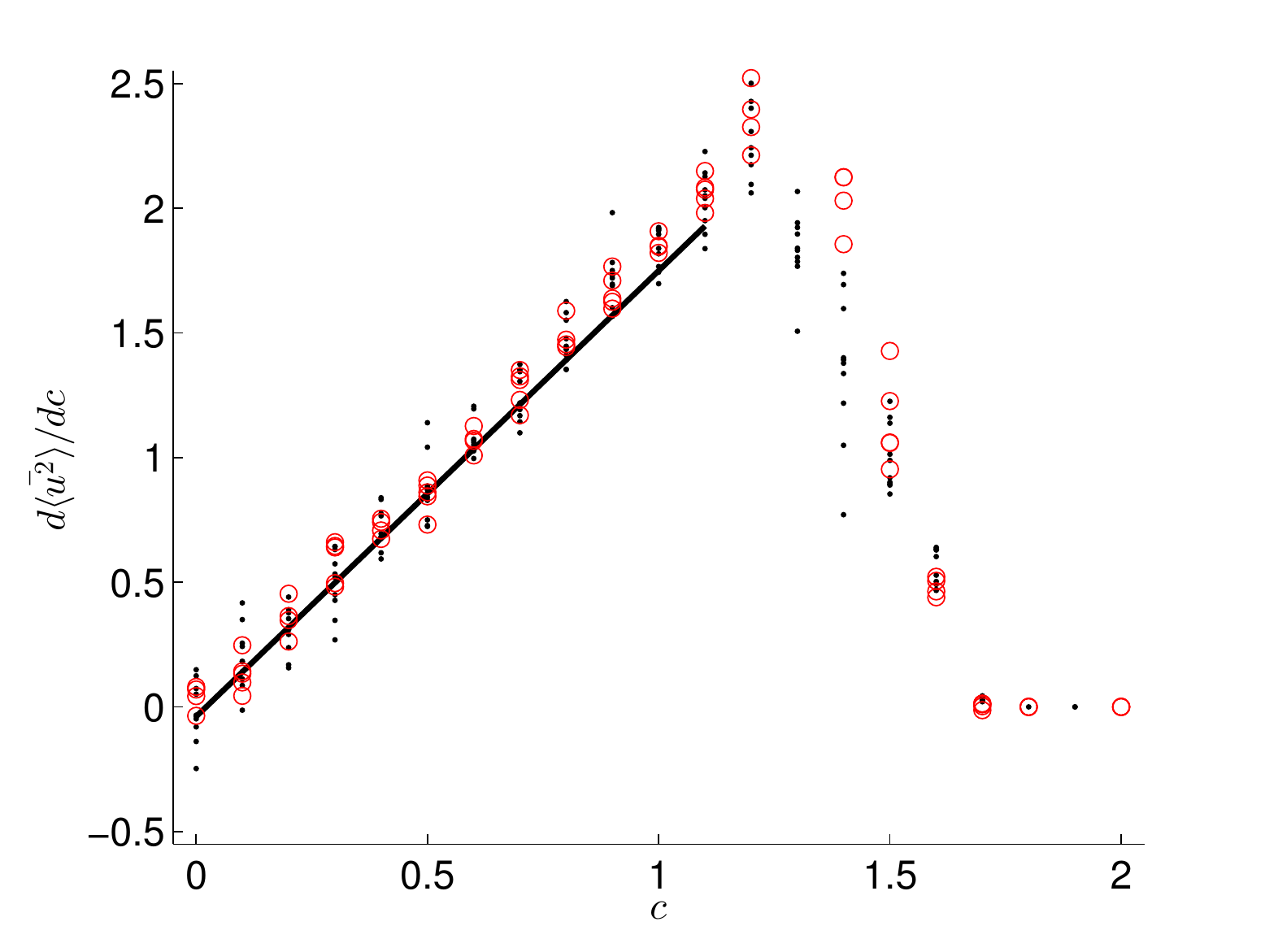}}
\caption{(a) $d\la \bar{u} \ra/dc$ versus $c$ and (b) $d\la \bar{u^2} \ra/dc$.  Each simulation was run for $T_0 = 1000$ time units before the LSS method was applied.  Black dots indicate gradients computed with a $100$ time unit intervals, the red circles indicate a $1000$ time unit interval and the blue diamonds indicate a $4000$ unit time interval respectively.  The black lines are slopes of linear or quadratic regressions from the data in figure \ref{f:obj_vs_c}. The three $\sigma$ error intervals, where $\sigma$ is the standard error of the slope, are indicated by black dotted lines. $u(x,t)$ was solved for with $n=511$ nodes, but LSS was conducted using $n=127$ nodes.  } 
\label{f:grad_vs_c}
\end{figure}

As discussed in section \ref{s:LSS}, we can use shadow trajectories to compute sensitivities of long-time averaged quantities to system parameters.  Figure \ref{f:grad_vs_c} shows this for $\la \bar{u} \ra$ versus $c$. The gradients computed using $100$ and $1000$ time unit intervals in the light turbulence dominated region ($0\le c \le 1.2$) are a good match for $\la J \ra =\la \bar{u^2} \ra$, but slightly over-predict the magnitude of the gradient for $\la J \ra= \la \bar{u} \ra$. The insensitivity of $\la\bar{u}\ra$ and $\la\bar{u^2}\ra$ to $c$ in the steady region $c \ge 1.8$ is also computed by the LSS method.  

However, LSS computes inaccurate gradients in the convection dominated regime ($1.2 < c \le 1.8$) for both objective functions considered. Possible reasons for this, and slight error in the sensitivity of $\la\bar{u}\ra$ are discussed in the next section.  

\subsection{Sources of error}
\label{ss:error}

One source of error is the difference between the infinite time shadowing direction and its approximation by LSS.  Our numerical shadowing direction $v_{num}(x,t)$ is just a finite time approximation of the infinitely long shadowing direction $v_{\infty}(x,t)$. It has been proven that the errors of this approximation are the greatest at the beginning ($t=0$) and end ($t=T$) of the numerical shadowing direction \cite{Wang:2013:LSSthm}.  Additionally, these errors decay exponentially in time.  The $t=0$ error decays forward in time at the rate of the smallest negative Lyapunov exponent. The $t=T$ error decays backward in time at the rate of the smallest positive Lyapunov exponent.  Therefore, $v_{num}(x,t)$ most accurately approximates $v_{\infty}(x,t)$ in the middle of the time interval it is computed on \cite{Wang:2013:LSSthm}. 

\begin{figure} 
\centering
\includegraphics[width=3.2in]{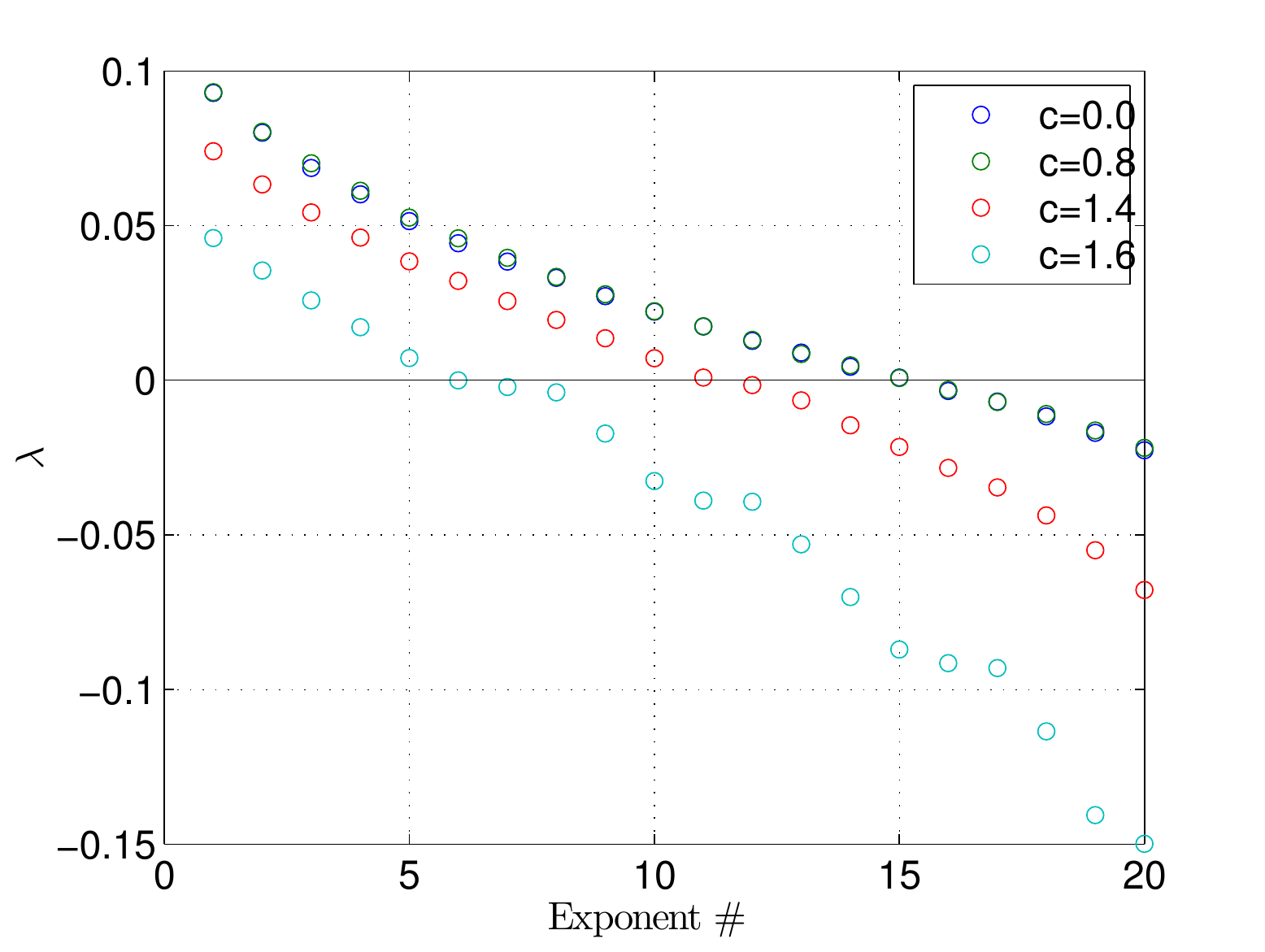}
\caption{Spectrum of Lyapunov exponents $\lambda$ for different values of $c$.  Note that the spectra for $c$ values in the convection dominated regime have two values very close to zero, as opposed to just one.  One of these corresponds to the neutrally stable exponent $\lambda = 0$, the other is just a very small magnitude exponent.  The exponents were computed using the method described by Benettin et al. \cite{Benettin:1980:Lyapunov}, with $s=100$ and $k=20000$.  Note that a time step size of $\Delta t = 0.05$ was used.  } 
\label{f:lyapunov}
\end{figure}

\begin{figure} 
\centering
\includegraphics[width=3.2in]{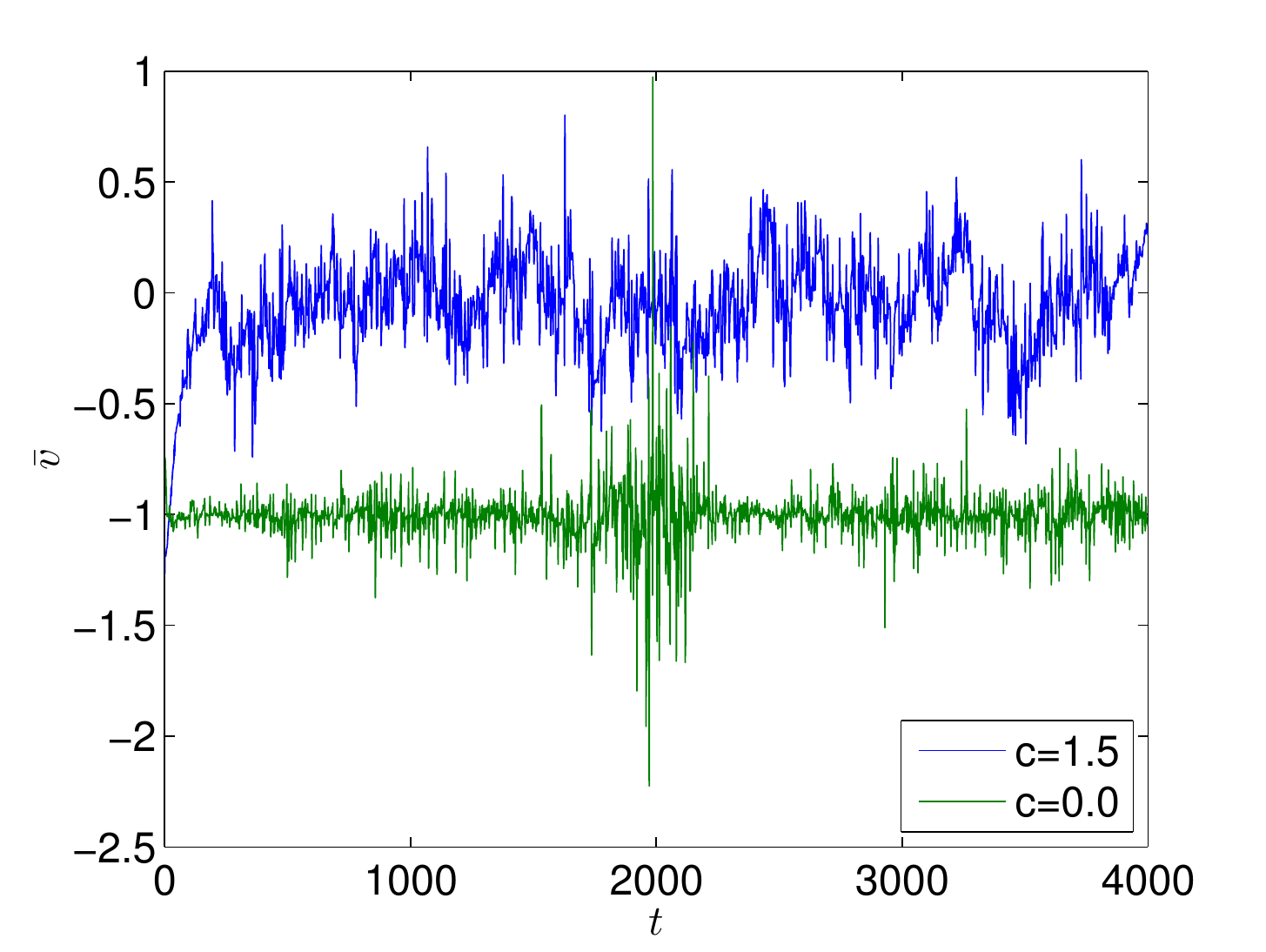}
\caption{Spatially averaged shadowing direction $\bar{v}$ versus $t$ for $c=0$ and $c=1.5$.  Each simulation was run for $T_0 = 1000$ time units before the LSS method was applied.  While the mean shadowing direction for $c=0$ converges to a quasi-steady state almost instantly, it takes until around $t=300$ for the $c=1.5$ shadowing direction to do so.  Finally, note the much larger magnitude oscillations and time scales associated with $c=1.5$. } 
\label{f:vbar_hist}
\end{figure}

This property of LSS suggests that the attractor associated with the convection dominated region has some Lyapunov exponents with magnitudes smaller than those in the light turbulence dominated regime.  This is shown to be the case in figure \ref{f:lyapunov}.  The low magnitude Lyapunov exponents cause the error in $v(x,t)$ for $c$ values in the convection dominated regime to decay very slowly as $t$ increases from $t=0$.  This slower convergence can be seen in figure \ref{f:vbar_hist}. 

Additionally, figure \ref{f:vbar_hist} shows that the slower time scale oscillations associated with the shadowing direction $v(x,t)$ are greater in magnitude in the convection dominated region than those present in the light turbulence region.  These longer time-scales require the use of a longer time interval to compute accurate gradients, as a low frequency periodic function would require a longer time interval to compute an accurate average.  The longer time scales observed in $v(x,t)$ in the convection dominated regime arise because of the smaller magnitude Lyapunov exponents, whose reciprocals are related to the time scales of $v(x,t)$.  

Overall, the convection dominated regime has a smaller magnitude non-zero Lyapunov exponent than the smallest non-zero Lyapunov exponent in the light turbulence dominated regime. This can contribute to the fact that the sensitivity computed over the same time interval is less accurate in the convection dominated regime.

However, even when a larger time interval is used, the gradients in the convection dominated regime appear to converge to the wrong value, as shown in figure \ref{f:grad_vs_c}.  Therefore, small magnitude Lyapunov exponents are not the only source of error.  To explore the other source of error present, we consider a different dynamical system, the sawtooth map:

\begin{figure} 
\centering
\includegraphics[width=2.0in]{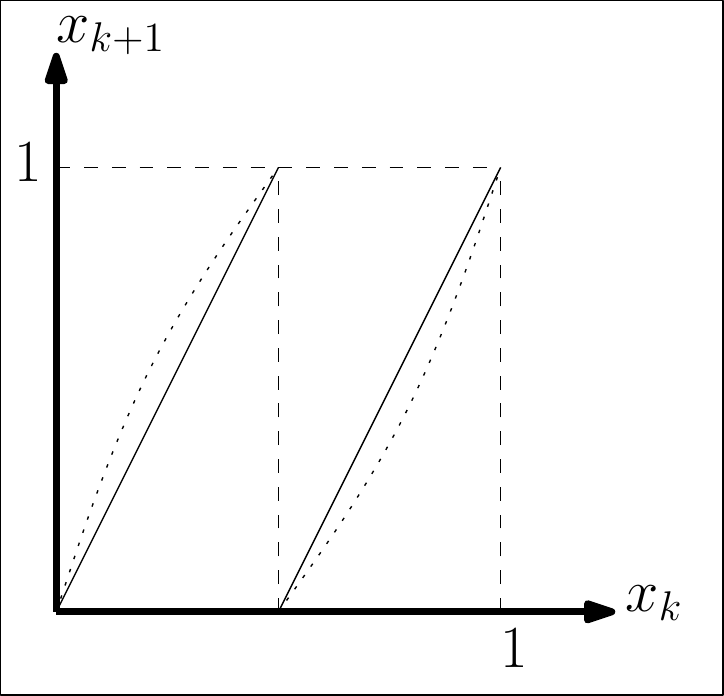}
\caption{Sawtooth map. The map for $s=0$ is indicated by the solid lines, the map for $s=0.1$ is indicated by the doted lines.  Dashed lines indicated the limits of phase space on the $x_k$ and $x_{k+1}$ axes.  } 
\label{f:sawtooth2}
\end{figure}

\begin{equation}
 x_{k+1} = F(x_k;s) = \left\lbrace\begin{array}{c}
2x_k + s\sin 2\pi x_k, \quad x_k \in [0,0.5] \\
2x_k - 1 + s\sin 2\pi x_k, \quad x_k \in (0.5,1.0]
\end{array} \right. 
\label{e:saw}
\end{equation}

\noindent For a visual representation of equation \eqref{e:saw}, refer to figure \ref{f:sawtooth2}.  The quantity of interest we consider for the sawtooth map is long time averaged $x^4$:

\begin{equation}
 \la x^4 \ra = \frac{1}{n} \sum_{k=1}^n x_k^4
\end{equation}

\begin{figure} 
\centering
\subfloat[Quantity of Interest]{
\includegraphics[width=3.2in]{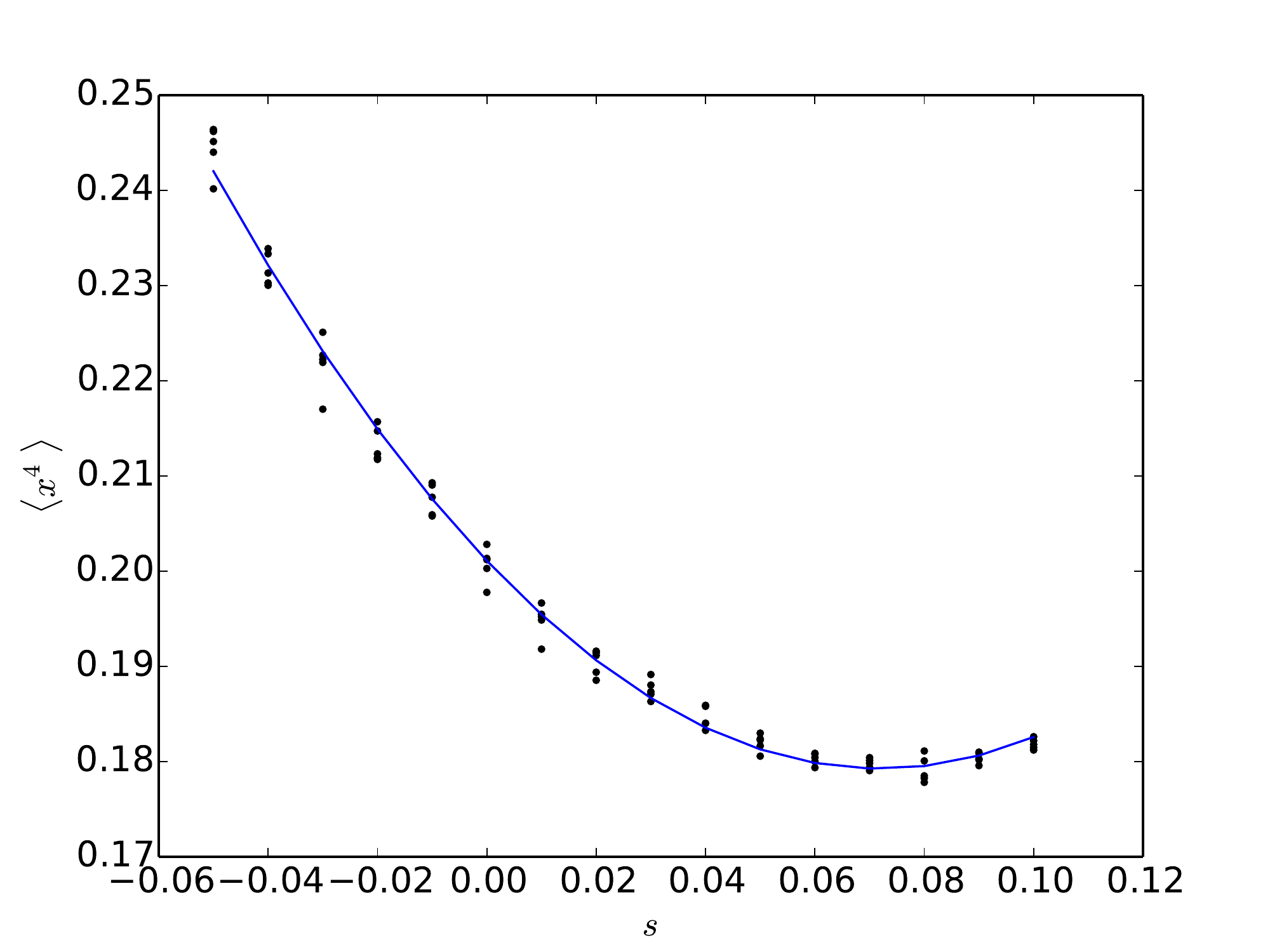}}\\
\subfloat[Gradients]{
\includegraphics[width=3.2in]{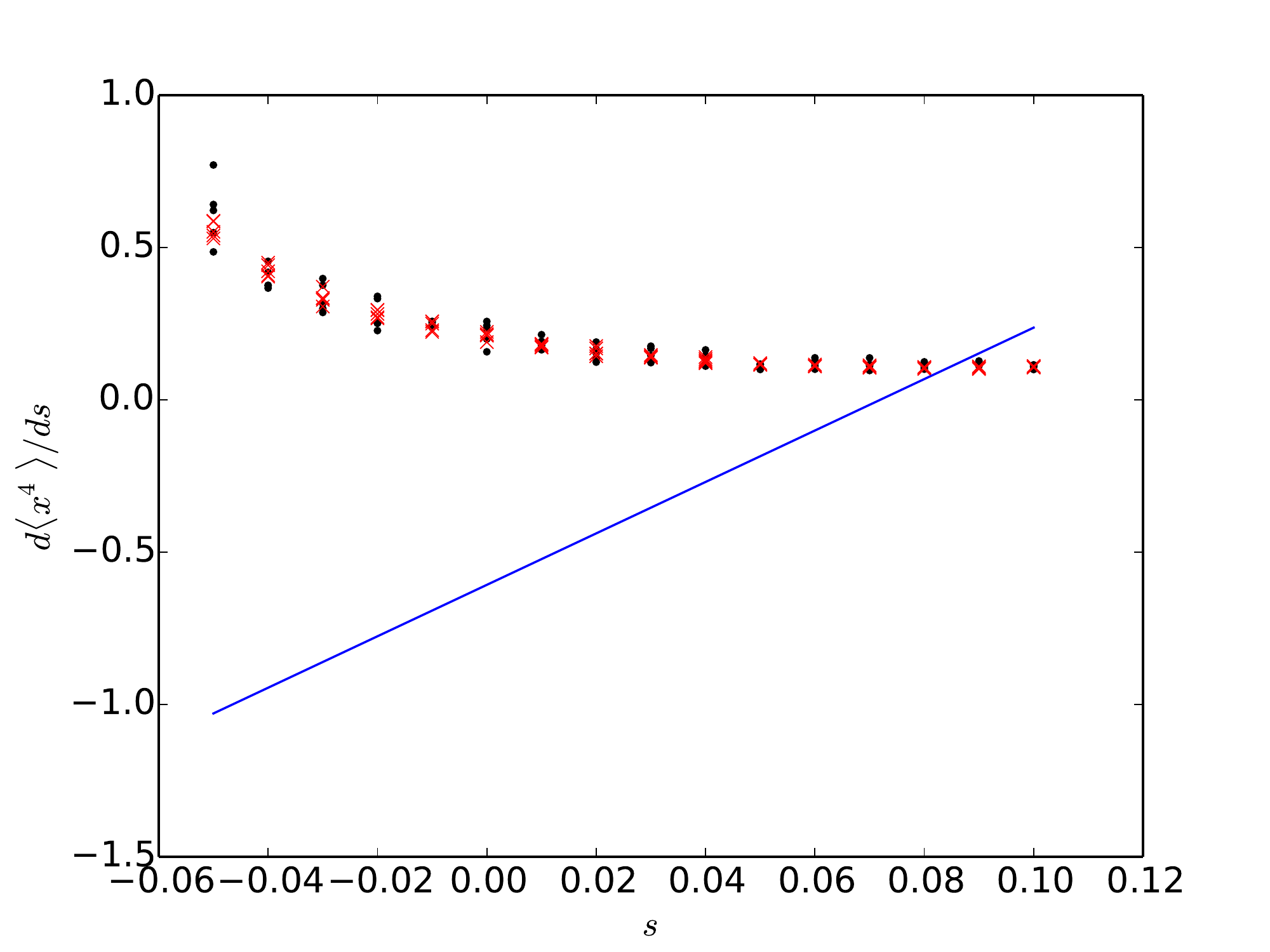}}
\caption{(a) $\la x^4 \ra$ realizations for difference $n=1\time 10^5$ length trajectories.  The solid line is a 2nd order polynomial curve fit is.   (b) $d\la x^4 \ra/ds$ realizations for $n=1\time 10^3$ and $n=1\time 10^4$ length trajectories in black and red, respectively.   The solid line is the first derivatives of the curve fit in (a).  } 
\label{f:saw_obj_grad}
\end{figure}

In figure \ref{f:saw_obj_grad}, we see that the gradients computed via LSS converge as $n$ is increased from 1000 to 10000, but they do not converge to the proper value, indicated by the slope of a 2nd order polynomial curve fit through the data in figure \ref{f:saw_obj_grad} (a).  This error is reminiscent of that observed for the K-S equation in the convection dominated regime.

%

To find the source of this error, we consider the stationary distribution of the sawtooth map.  The stationary distribution, $\rho(x;s)$, is a distribution in phase space, or $x \in [0,1]$ for the sawtooth map.  The value of $\rho(x;s)$ corresponds to how often a trajectory $\{x_1,x_2,...,x_n\}$ passes through $x$, for a given value of $s$.  It can be shown that time averaged quantities can be written as averages in phase space using the stationary distribution \cite{Blonigan:2013:pdf}, for instance:

\begin{equation}
 \la x^4 \ra = \frac{1}{n} \sum_{k=1}^n x_k^4 = \int_0^1 x^4 \rho(x;s) dx
 \label{e:stat_dist}
\end{equation}

\noindent From equation \eqref{e:stat_dist}, we can see that variations in $\la x^4 \ra$ with $s$ can be expressed as variations of $\rho(x)$ with $s$.  Therefore, sensitivities of $\la x^4 \ra$ can be expressed in terms of sensitivities of the distribution \cite{Blonigan:2013:pdf}:

\begin{equation}
 \frac{d \la x^4 \ra}{ds} = \lim_{\Delta s \to 0} \left(\int_0^1 x^4\frac{(\rho(x;s+\Delta s)-\rho(x;s))}{\Delta s} dx \right) = \int_0^1 x^4 \pdI{\rho}{s} dx
 \label{e:stat_dist_deriv}
\end{equation}

\noindent We compute $\rho_s(x)$ empirically with the following algorithm:

\begin{enumerate}
 \item Compute a very long trajectory $\{x_1,x_2,...,x_n\}$.  $n=1\times 10^6$ is sufficient for the sawtooth map.  
 \item Divide phase space into equally sized intervals.  40 were used for the results presented in this paper.  
 \item Compute the frequency that $\{x_1,x_2,...,x_n\}$ falls in each interval and normalize it with $n$ to form a histogram.  
\end{enumerate}

\begin{figure} 
\centering
\includegraphics[width=3.2in]{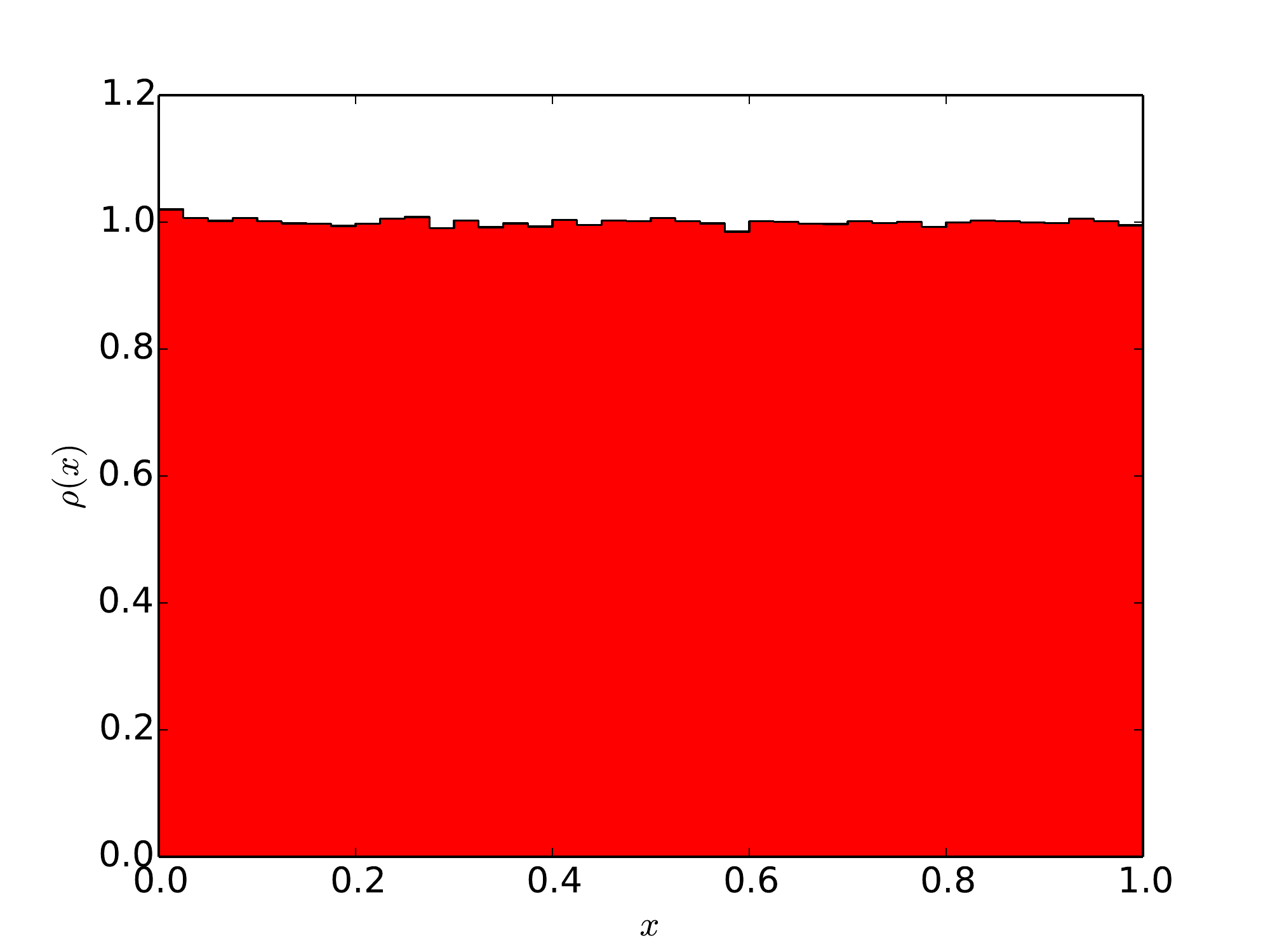}
\caption{Typical empirical stationary distribution for the sawtooth map with $s=0$.  The distribution was formed using a $n=1\times 10^6$ length sequence of $x_k$'s.   } 
\label{f:sd_s0}
\end{figure}

\begin{figure} 
\centering
\subfloat[Reference Trajectory, $s=0.1$]{
\includegraphics[width=3.2in]{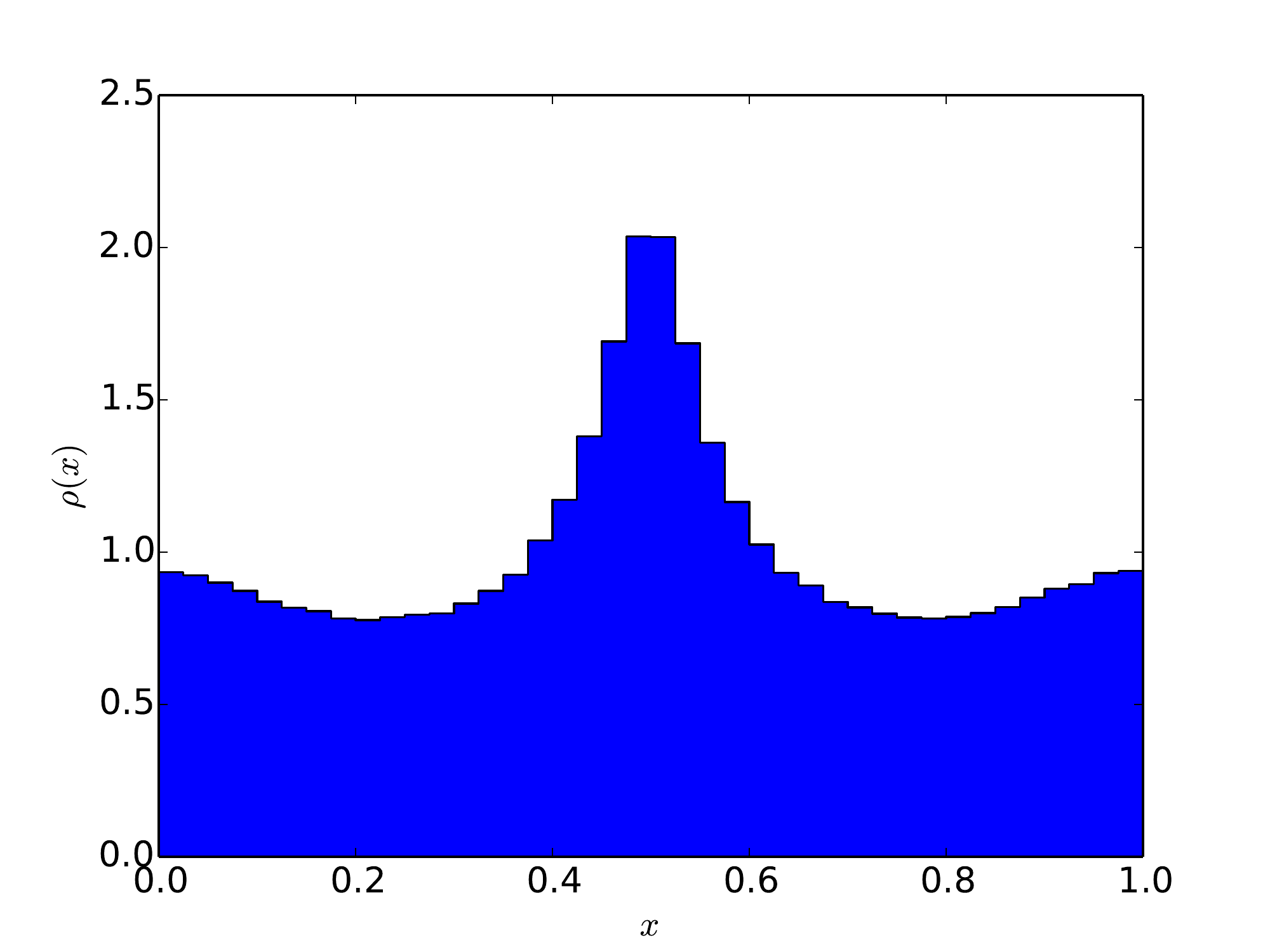}}\\
\subfloat[Shadow Trajectory, $s=0.0$]{
\includegraphics[width=3.2in]{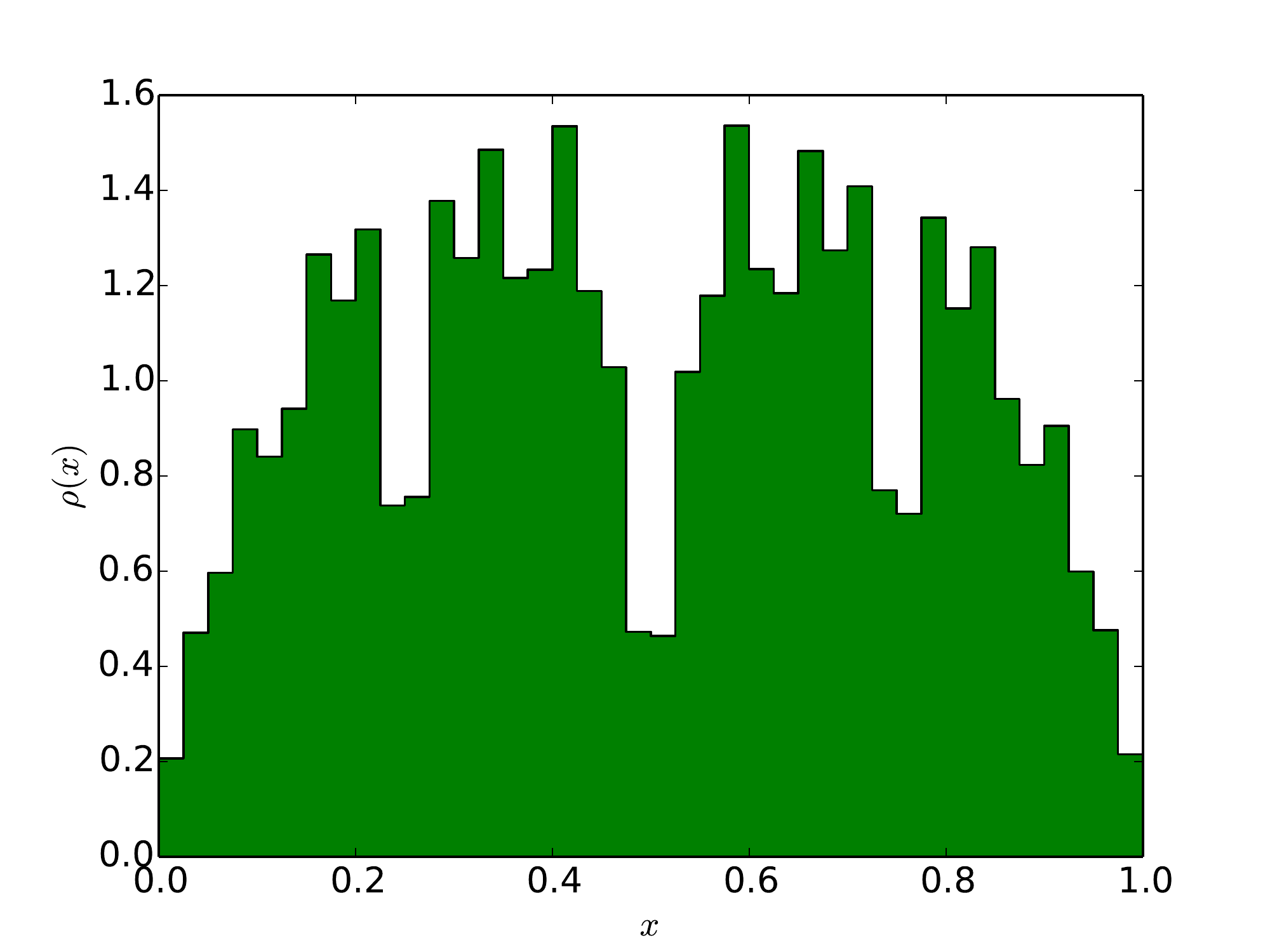}}
\caption{Typical empirical stationary distribution for the sawtooth map with $s=0.1$ and the stationary distribution for the corresponding shadow trajectory with $s=0.0$.  Both distributions were formed using a $n=1\times 10^6$ length sequence of $x_k$'s.   } 
\label{f:sd_ref_sha}
\end{figure}

For the case $s=0$, figure \ref{f:sd_s0} shows that the stationary distribution is approximately uniform ($\rho(x) = 1$).  From figure \ref{f:sd_ref_sha} a), we see that the distribution changes as $s$ is increased.  Now, when we compute the $s=0$ shadow trajectory of the $s=0.1$ trajectory shown in figure \ref{f:sd_ref_sha} a), we expect to see a uniform distribution, as in figure \ref{f:sd_s0}.  However, this is not the case.

From figure \ref{f:sd_ref_sha} b), we see the stationary distribution $\rho(x)$ of the shadow trajectory is not uniform at all.  Therefore, the expression for the derivative in equation \eqref{e:stat_dist_deriv} is incorrect, as observed in figure \ref{f:saw_obj_grad}.  In other words, a shadow trajectory with $s=s_1$ of some reference trajectory with $s=s_0$ does not necessarily have the same long time averaged quantities as a typical trajectory with $s=s_1$.  

Obtaining different long time averaged quantities from solutions with the same values of $s$ (or $c$ for the K-S equation) contradicts the assumption of ergodicity.  It can be shown that LSS will compute correct gradients if the long time averaged quantities are the same for all initial conditions \cite{Wang:2013:LSSthm}.  However, in practice long time averaged quantities are only the same for {\it almost all} initial conditions.  For example, the long time averaged quantities of the K-S equation are all zero for the initial condition $u(x,0) = 0$ when $c=0.5$, which is different from the values in figure \ref{f:obj_vs_c}.  Also, if we consider a trajectory of the sawtooth map starting at $x=1.0$ will stay at $x=1.0$ for all iterations, resulting in $\la x^4 \ra = 1$, regardless of the value of $s$, in contrast to the trend shown in figure \ref{f:saw_obj_grad}.  Both the K-S equation and the sawtooth map are not strongly ergodic.  That is, the same long time averaged quantities are obtained for {\it almost all}, not {\it all} initial conditions, so computation of the correct gradient is not guaranteed.  

It is important to emphasize that LSS has been used to compute very accurate gradients for a number of smaller systems, including the Lorenz system \cite{Wang:2013:LSS2}, and that it computes fairly accurate gradients for the K-S equation when $|c| \le 1.2$ and $|c| \ge 1.7$.  Therefore, it seems that this breakdown in LSS occurs in largely varying degrees.  Further work needs to be done to determine what specific properties of the sawtooth map cause this issue to occur.

\section{Conclusion}
\label{s:conclusion}

In conclusion, the LSS method computes accurate gradients for a wide range of system parameter values for a modified K-S equation. In particular, it was found that LSS worked very well in the light turbulence dominated regime, especially for the quantity of interest $\la \bar{u^2} \ra$.  However, the method slightly over-predicted the magnitude of gradients of the quantity $\la \bar{u} \ra$ in the light turbulence regime and falsely predicted all sensitivities in the convection dominated regime.  This breakdown in the method is also observed for smaller systems, such as the sawtooth map.  It appears that while LSS works very well for some systems, including the Lorenz system and the K-S equation for certain values of $c$, it does not work very well or at all for other systems. 

Future work needs to be done to determine what properties of the sawtooth map and others like it cause LSS to break down.  The lessons learned from working with the sawtooth map can potentially be applied to making LSS more robust and learning what classes of PDEs and larger systems LSS is best suited for.

\section*{Acknowledgements}
The authors would like to acknowledge AFSOR Award F11B-T06-0007 under Dr. Fariba Fahroo, NASA Award NNH11ZEA001N under Dr. Harold Atkins as well as financial support from the NDSEG fellowship.  

\section*{References}

\bibliographystyle{elsarticle-num}

\end{document}